\documentclass[journal]{IEEEtran}
\ifCLASSINFOpdf
\else
\fi

\usepackage[numbers]{natbib}
\usepackage{amssymb}
\usepackage{graphicx}
\usepackage{amsmath,amssymb,amsfonts}
\usepackage{algorithmic}
\usepackage{textcomp}
\usepackage{wrapfig}
\usepackage[utf8]{inputenc}
\usepackage[T1]{fontenc}
\usepackage{amsmath}          
\usepackage[bottom]{footmisc}
\usepackage{amssymb}          
\usepackage{amsthm}
\usepackage{subfigure}
\usepackage{subcaption}
\usepackage{epstopdf}
\usepackage{array}
\usepackage{multirow}
\usepackage[
    binary-units=true,
    range-units=single,
    list-units=single,
    separate-uncertainty=true,
    per-mode=symbol,
    ]{siunitx}
\usepackage{color}
\usepackage{balance}

\usepackage[table,xcdraw]{xcolor}
\usepackage{booktabs}
\usepackage{diagbox}

\usepackage{makeidx}  
\usepackage{url}
\usepackage{bm}
\usepackage{tabularx}
\usepackage{array}
\usepackage[justification=centering]{caption}
\usepackage{algorithm}
\usepackage{algorithmic}
\usepackage{graphicx}
\usepackage{makecell}
\usepackage{subcaption}
\usepackage{siunitx}

\begin{document}
%
\title{Topology Partitioning-based Self-Organized Localization  in Indoor WSNs with Unknown Obstacles}
%

%

        
 \author{Ze Zhang, Qian Dong$^{*}$

\thanks{Corresponding author: Qian Dong (Email: qian.dong@xjtlu.edu.cn)}

\thanks{Z. Zhang and Q. Dong are with the School of Advanced Technology, Xi’an Jiaotong-Liverpool University, China. Email: Ze.Zhang20@student.xjtlu.edu.cn}





}

\maketitle

\begin{abstract}
Accurate indoor node localization is critical for practical Wireless Sensor Network (WSN) applications, as Global Positioning System (GPS) fails to provide reliable Line-of-Sight (LoS) conditions in most indoor environments. Real-world localization scenarios often involve unknown obstacles with unpredictable shapes, sizes, quantities, and layouts. These obstacles introduce significant deviations in measured distances between sensor nodes when communication links traverse them, severely compromising localization accuracy. To address this challenge, this paper proposes a robust range-based localization method that strategically identifies and severs obstructed communication paths, leveraging network topology to mitigate obstacle-induced errors. Across diverse obstacle configurations and node densities, the algorithm successfully severed $87\%$ of obstacle-affected paths on average. Under the assumption that Received Signal Strength Indicator (RSSI) provides accurate distance measurements under LoS conditions, the achieved localization accuracy exceeds $99.99\%$.

\end{abstract}


%
\IEEEpeerreviewmaketitle

\section{Introduction}

\IEEEPARstart{W}{ireless} sensor networks consist of numerous sensor nodes that collaborate with each other in an ad-hoc pattern to gather information through wireless signal transmission and multi-hop communication. Due to the large number and low cost of sensor nodes, WSNs are often deployed in environments where wired networks are either inaccessible, impractical or expensive. This has led to their widespread adoption across various domains, including the battlefield surveillance in military applications \cite{7762095}, smart home automation and factory production management within the Internet of Things (IoTs) framework\cite{6827681}\cite{22062087}, intelligent transportation in urban governance\cite{9319346} \cite{5941955}, remote healthcare monitoring in the medical field\cite{5570866}, and disaster predictions and rescue operations in the emergency scenarios \cite{9697236}\cite{9301283}.

In order to achieve precise monitoring, accurate node positioning is crucial in most applications, especially for human-centric scenarios. However, due to the random distribution of nodes and the deployment of nodes in dangerous and remote areas without human intervention and infrastructure assistance, sensor nodes cannot predict or locate themselves without mounting positioning devices. This makes all the nodes are location-unknown terminals. But even if they are equipped with devices, like carrying GPS components which has been considered to be a feasible solution in exterior cases, are insufficient to accurately localize sensor nodes in indoor environments, where the LoS condition is unsatisfied \cite{7762095}. Therefore, pre-deploying nodes in predetermined locations is a more practical solution for indoor localization. These nodes with known coordinates are considered anchor nodes. By taking utilization of anchor nodes, most of the nodes in the network which are unknown of their coordinates are able to locate themselves. This is deemed more operationally feasible\cite{10.1007/s11277-021-08945-8}.

To determine the locations of unknown nodes based on anchor nodes, either range-free or range-based localization algorithms can be employed. Unlike range-free methods, which rely solely on the connectivity relationships among nodes, range-based methods require the measurement of distances or angles between nodes. Given that our WSN has a limited number of anchor nodes and a large number of unknown nodes randomly deployed across a wide area, the range-based algorithm provides a distinct advantage by delivering more accurate evaluation results. Since the Received Signal Strength Indicator (RSSI) is an embedded technology in sensor nodes that does not require additional hardware or incur extra costs, it is widely adopted for estimating the relative distances between nodes in WSNs \cite{10.5296/npa.v2i1.279}.

Although the RSSI technique has been proven effective for estimating relative distances between nodes in LoS scenarios, its accuracy degrades significantly in real indoor environments due to the presence of obstacles, which introduce Non-Line-of-Sight (NLoS) effects \cite{CARPI2023109663}. In such environments, convex corners of obstacles between transceiver pairs can cause complex signal propagation phenomena, including reflection, refraction, diffraction, and scattering, leading to severe signal decay. Additionally, the attenuation rates of electromagnetic waves differ between air and obstacles, resulting in significant discrepancies between the received signal strength and its expected value, further distorting RSSI measurements for distance estimation when signals traverse obstacles. Even worse, obstacles can block communication between transceivers, leading to errors in estimating the minimum hop count between two multi-hop nodes. In such cases, summing the estimated distances of multiple one-hop links to approximate the total distance between these nodes will result in overestimation, further exacerbating localization inaccuracies. These challenges highlight the limitations of RSSI-based localization in complex indoor environments, where the number, shape, size, locations, and deployment patterns of obstacles are unknown. Consequently, RSSI measurements affected by obstacles cannot be reliably used for distance estimation in indoor WSNs \cite{10.1063/5.0160313}.

To address the impact of obstacles on distance estimation between nodes, this paper proposes partitioning the entire network into distinct sub-networks. This ensures that each divided sub-network is free of convex obstacle corners, allowing the localization of unknown nodes to be conducted within each sub-network independently. By adopting this strategy, the adverse effects of obstacles on the localization accuracy of unknown nodes can be completely eliminated \cite{10.1145/1161089.1161104, PHOEMPHON2024103783, 7000585, 9684470}.

Therefore, to accurately determine the coordinates of unknown nodes in large-scale WSNs with unknown obstacles, this paper proposes a three-step approach: (1) identifying segmentation nodes at convex obstacle corners, (2) partitioning the network to convert convex obstacles into concave ones, and (3) localizing all unknown nodes using a unique relative coordinate system in each sub-network.

Specifically, assuming each obstacle edge exceeds the nodes' maximum radio range, communication between nodes on opposite sides of a convex corner is blocked, forcing their shortest paths to traverse the corner. When computing the shortest paths between any two nodes, the nodes that appear most frequently in these paths are identified as segmentation nodes for each convex obstacle corner. Based on the segmentation node pairs, bisectors at the exterior angle of convex obstacle corners are formed by nodes equidistant to the two segmentation nodes in each pair, dividing the network into sub-networks free of convex corners. Nodes are then assigned to their designated sub-networks based on hop count differences to the segmentation nodes in each pair. Within each sub-network, due to the limited number of anchor nodes and the overestimation of distances when unknown nodes are more than one hop away from anchor nodes, a unique relative coordinate system is established. This system is built from three reference nodes, which comprise a mix of unknown and anchor nodes, are within one hop of each other, non-collinear, and centrally positioned. The relative coordinates of unknown nodes are determined based on their proximity to one, two, or all reference nodes in each sub-network. Finally, these relative coordinate systems are calibrated to the global coordinate system, enabling accurate localization of all unknown nodes across the network.

\begin{figure}[htbp]
	\centering
	\subfigcapskip=-15pt
	\subfigure[The original sender seizes the channel]{
		\begin{minipage}[b]{0.45\textwidth}
			\includegraphics[width=1\textwidth]{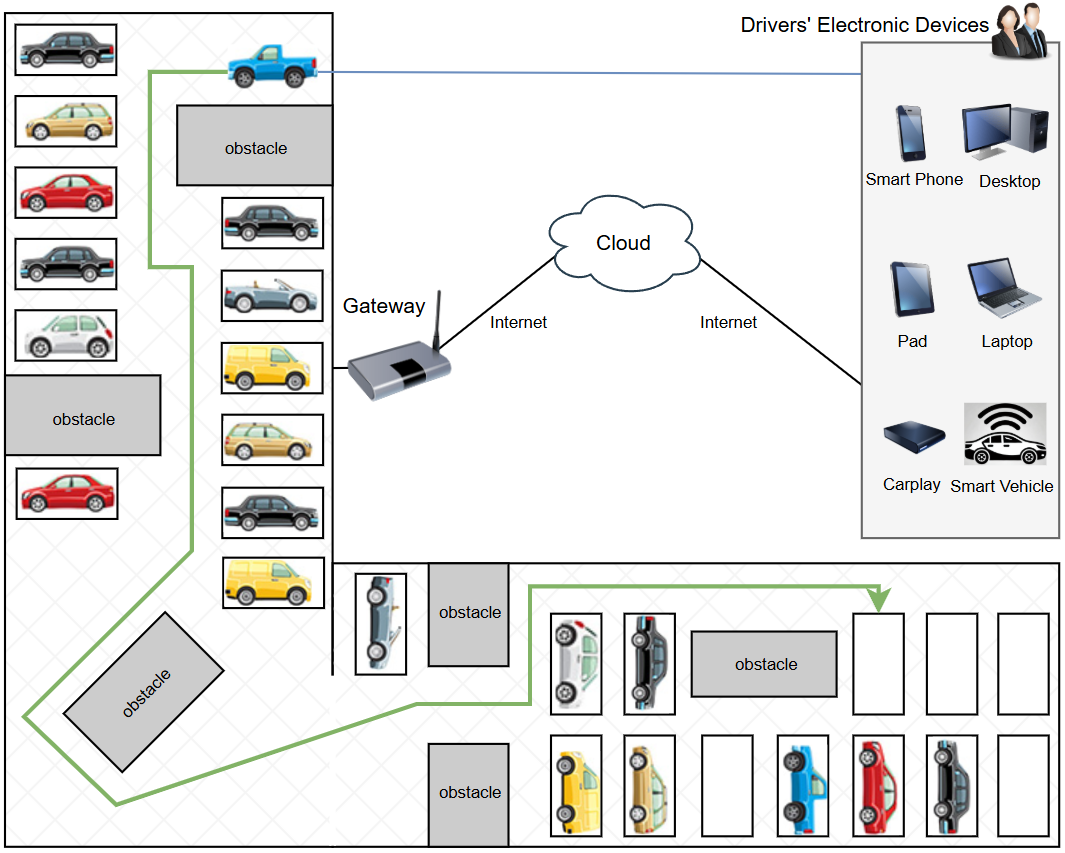} \\
		\end{minipage}
	}
	\subfigure[Another sender seizes the channel]{
		\begin{minipage}[b]{0.45\textwidth}
			\includegraphics[width=1\textwidth]{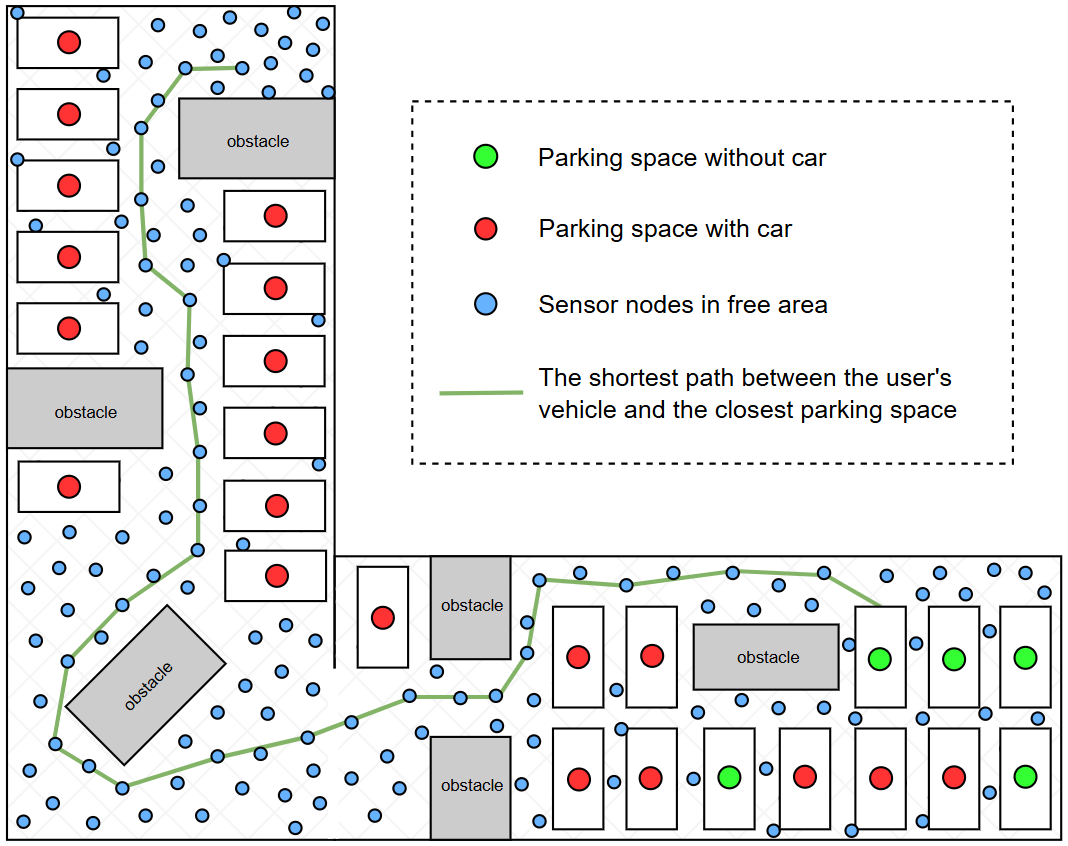} \\
		\end{minipage}
	}
	\caption{Our design can be applied to autonomous driving in indoor garages, where the WSN is deployed to detect the shape and position of obstacles, localizes sensor nodes, and dynamically determines the shortest path to the nearest available parking space. This enables real-time map construction and precise navigation in indoor environments.}
	\label{f}
\end{figure}

Our design can assist in autonomous driving applications, where available parking spaces are typically displayed when a self-driving car enters an indoor garage. However, due to unfamiliarity with the layout of shopping mall or office building garages and the absence of fixed parking spots, self-driving cars often only know the general direction of available spaces without being able to pinpoint their exact locations. To address this issue, this paper proposes a two-step solution. First, it dynamically evaluates the positions and shapes of obstacles, enabling the system to calculate the shortest path from the car's current location to the nearest available parking space in real time, thereby avoiding traffic congestion and improving parking efficiency. Second, it precisely localizes all nodes in the parking lot, allowing the system to compute the real-time distance between the vehicle and the nodes along the shortest path. This guides the vehicle from the garage entrance to the nearest available parking space, enabling map construction and navigation functionality for parking services in dynamic indoor environments.

As shown in Figures \ref{f}(a) and \ref{f}(b), a partial area of an indoor garage is depicted, where sensor nodes are randomly deployed throughout the network. Additionally, each parking space is equipped with a sensor node, with red and green lights indicating whether the space is occupied or available. Various obstacles exist within the garage, including load-bearing walls, moving vehicles, improperly parked cars, and shopping carts. Our system continuously updates the real-time shortest path to the nearest available parking space based on obstacle conditions. Once this information is processed, the WSN interacts with the 5G network via a gateway, enabling self-driving cars to respond accordingly.

Aiming at designing a topology partitioning-based localization for indoor WSNs with unknown obstacles, the proposed method reduces system complexity, minimizes reliance on anchor nodes, adapts to various obstacle conditions, and provides accurate positioning of unknown nodes. The key contributions are:

\begin{itemize}
\item[$\bullet$] \textbf{Identification of segmentation node pairs for convex obstacle corners.} The shortest paths between any two nodes are analyzed, and the nodes that appear most frequently in these paths are identified as segmentation nodes. This enables the precise detection of convex obstacle corners in WSNs.
\item[$\bullet$] \textbf{Network partitioning using bisectors to transform obstacles from convex to concave.} Bisectors are constructed based on segmentation node pairs, partitioning the network into sub-networks free from convex obstacles. This eliminates obstacle-induced errors in distance estimation in each sub-network. This method relies solely on node connectivity and is independent of anchor nodes, ensuring greater generality and adaptability in network partitioning, regardless of the shape, position, quantity, size, and layout of obstacles.
\item[$\bullet$] \textbf{Accurate distributed localization using relative coordinate systems in each sub-network.} The method relies solely on network topology to establish and calibrate the relative coordinate system, enabling localized position estimation. This mitigates multi-hop distance estimation errors, reduces reliance on global anchor nodes, and enhances the localization robustness and scalability. Its distributed evaluation aligns with the decentralized nature of WSNs and its parameter-free design eliminates the need for scenario-specific tuning, offering greater adaptability, universality, and scalability for real-world localization challenges.
\end{itemize}

The rest of this paper is organized as follows. Section II discusses the impact of convex corners of obstacles on RSSI-based distance evaluation. Section III identifies segmentation nodes based on their frequency of occurrence in shortest paths between any nodes in the network. Section IV determines segmentation node pairs, constructs bisectors for each pair, and partitions the network into sub-networks. Section V establishes a relative coordinate system for each sub-network, localizes unknown nodes within each sub-network, and calibrates the relative coordinate system to the global coordinate system. Section VI compares analytical and simulation results regarding partition effectiveness and localization accuracy. Section VII concludes this paper.

\section{ Distance Evaluation with Obstacles}

The RSSI technique, which is initially integrated into sensor nodes, is able to assist nodes in estimating the relative distance between each other without the need for extra device and cost, thereby, it is widely used in WSNs for positioning. However, the RSSI value will be inaccurate when an obstacle stand between the transceiver. This NLoS phenomenon can lead to severe errors in localization. In this section, this paper will introduce RSSI technology, the obstacle categories and characteristics, as well as the impact of obstacles on RSSI values, which acts as the basis for algorithm design in subsequent sections.

\subsection{RSSI Errors in Localization caused by Obstacles}

Errors in RSSI measurement arise from various factors. A significant and mathematically challenging contributor is the reflection, refraction, diffraction, and scattering of signals caused by obstacles in indoor localization environments. Additionally, the attenuation rates of electromagnetic waves differ between air and obstacles, leading to further inaccuracies in the distance measured by RSSI for nodes situated near obstacles. Moreover, the dramatic signal decay caused by obstacles and the energy absorbed by them result in the considerable fluctuations in the received signal values. Consequently, the RSSI values and the relative distance between nodes will not present a one-to-one relationship when obstacles positioned in the middle. As a result, the received signal strength will differ significantly from the expected actual value it should be. This makes the received signals which traverse obstacles cannot be used as criteria for distance evaluation.

\begin{figure}[htbp]
	\centerline{\includegraphics[width=2in]{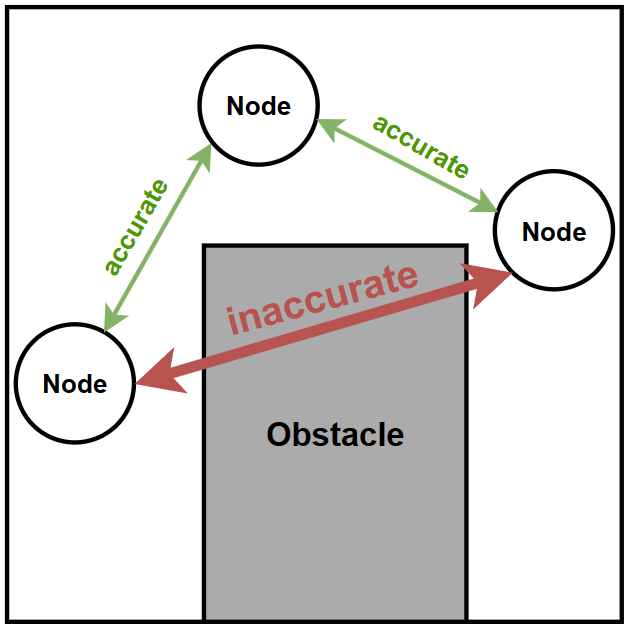}}
	\caption{In RSSI technique, obstacles attenuate the RSSI signals between two nodes, leading to inaccurate distance estimation where errors are highly increased.}
	\label{1}
\end{figure}

To address this issue, we leverage the connectivity relationships among WSN nodes and implement partitioning strategies to minimize the impact of obstacles on the accuracy of RSSI measurements. The inaccuracy localization caused by obstacles is illustrated in Figure \ref{1}.

\subsection{Convex and Concave Corners of Obstacles}

\begin{figure}[htbp]
\centerline{\includegraphics[width=3in]{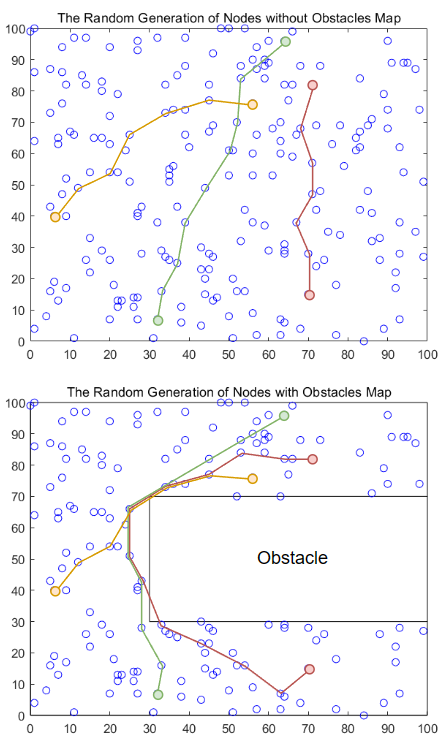}}
\caption{The convex obstacles have the potential to augment the number of hops on shortest path between two nodes. The inaccurate estimation of the distance increases the error in node localization.}
\label{2}
\end{figure}

The obstacles located within the WSNs cannot be disregarded, as they are not only impact the accuracy of distance measurements, but also cause errors in the estimation of the minimum number of hops between multihop nodes \cite{4205060}. As shown in Figure \ref{2}, if no obstacles exist in the network, the shortest path between any two nodes approximates a straight line, as the yellow, red, and green lines shown by the top sub-figure. However, when an obstacle starts to appear, the shortest path between part of nodes must navigate around it, resulting in a more curved route, as shown in the bottom sub-figure. In this case, using the summation of each one-hop distance to estimate the relative distance between any two nodes will significantly exceed their actual distance, leading to localization inaccuracy.

To distinguish whether obstacles have convex and concave corners, this paper references the definitions of convex and concave functions originally defined in mathematics. In the context of differentiable functions, $f$, is considered to be a convex function if and only if, $\forall x,y \in dom f$, the following inequality holds:

\begin{equation}
f(y) \geq f(x) + \nabla f(x)^T (y-x)
\end{equation}

 In Inequality 5, terms $x$ and $y$ are any points in the domain of function $f$, while $f (x)$ and $f (y)$ represent the values of the function in $x$ and $y$. Term $\nabla f(x)$ represents the gradient of the function at point $x$. This inequality indicates that the tangent line at a point  $x$ (given by the gradient $\nabla f(x)$) is lower than or equal to the value of the function $f$ at any other point $y$. This means that for any point $y$, the function value $f(y)$ is always greater than or equal to the value of the tangent line at point $x$. In simpler terms, this inequality shows that if point 
 $x$ is moved to point $y$, the value of the function $f$ will not fall below the tangent line value at point $x$. This property is a key characteristic of convex functions, indicating that the function curves "upward" in that interval, ensuring that there are no local depressions. The definition of a concave function is the opposite of that of a convex function, as expressed by Inequality 6.

 \begin{equation}
 	f(y) \leq f(x) + \nabla f(x)^T (y-x)
 \end{equation}

However, obstacles may have irregular surfaces, such as protruding sharp corners. In such cases, function $f$ is not differentiable, and the aforementioned inequalities cannot be used for determination of convex and concave corners of obstacles. Additionally, the definitions of concave and convex functions are relative, as the concavity or convexity of the same obstacle may vary when observed from different coordinate systems or perspectives.

Therefore, we define the concave and convex corners of obstacles in a two-dimensional area using another method. In a region on the boundary of an obstacle:
\begin{itemize} 
	\item[$\bullet$]If the line segment connecting the two arbitrarily chosen points lies entirely within the interior of the obstacle, \textbf{the obstacle has a convex corner} in the interval between these two points.
    \item[$\bullet$] However, when the line segment connecting these two points lies entirely outside the obstacle, \textbf{the obstacle has a concave corner} in the interval between these two points.
    \item[$\bullet$] In addition to these two cases, when the line segment connecting the two points coincides with the boundary of the obstacle, \textbf{the boundary of the obstacle is a straight line} between these two points. 	 
\end{itemize}
 	 
\begin{figure}[htbp]
\centerline{\includegraphics[width=3.5in]{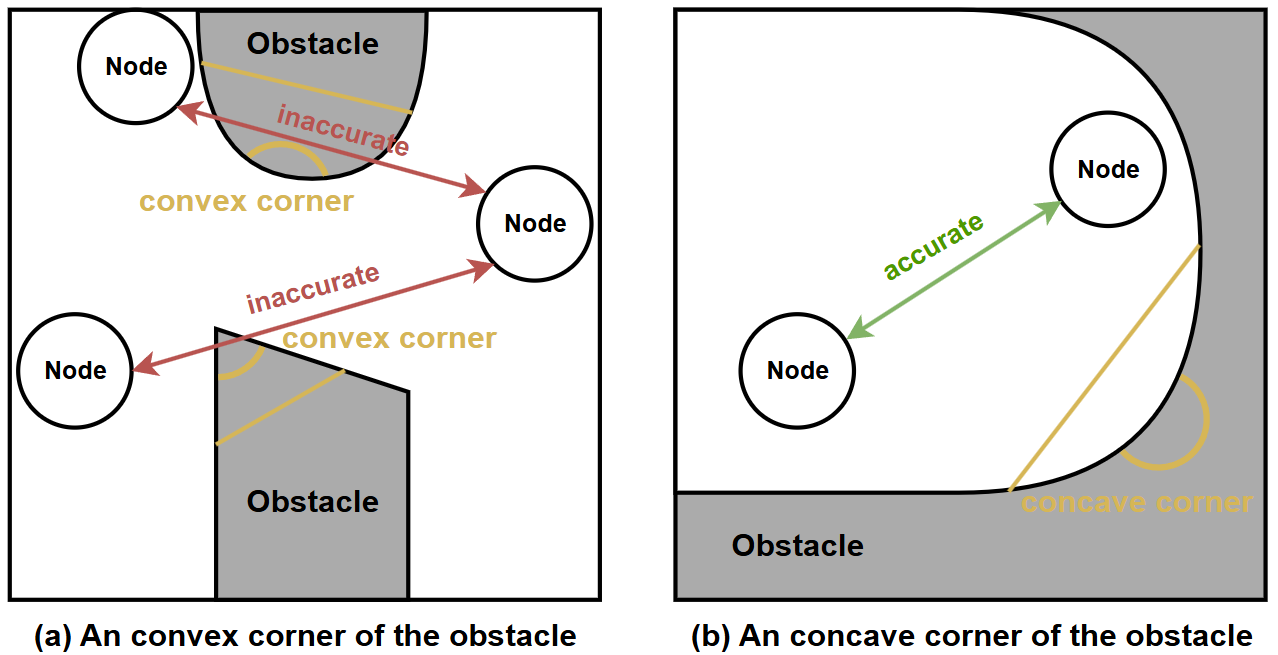}}
\caption{The convex and concave corners of obstacles and their impact on the received signal quality}
\label{3}
\end{figure}

 For the communication between two nodes surrounding an convex corner of an obstacle, their signal transmission will be blocked, resulting in inaccurate measurements of relative distance between them.  On the contrary, the existence of concave corners of obstacles can be viewed as ensuring uninterrupted communication between nodes. The impact of convex corner of obstacle on the receiving signal quality is illustrated in Figure \ref{3}(a), while the unobstructed communication link in case of concave obstacle is depicted in Figure \ref{3}(b).

\section{Determination of Segmentation Nodes }

This paper assumes that the length of each edge of the obstacle is larger than the maximum radio transmission range of nodes. Under this assumption, the shortest path for communication between a pair of nodes located on both sides of the protruding corner of the obstacle will be blocked. This will change the shortest path between these two nodes, as they have to traverse the convex corner of the obstacle. If the shortest path between any two nodes in the network is counted, there must emerge certain characteristic nodes, which are typically situated near the convex corner of the obstacle. 

If the distribution of nodes is not extremely uneven, we can definitely find at least one pair of characteristic nodes that appear much more frequently on the shortest path between any two nodes in the network than other nodes in the vicinity of the convex corner of an obstacle. 
 After each characteristic node is identified, these nodes can serve as the foundation for the network partitioning which divides the entire network into segments, so that there will be no convex corner of obstacles anymore in each of the network segmentation. This ensures that the communication between nodes in each sub-network will not be interrupted.

This section uses the H-shaped obstacle as a case study to determine the shortest paths between every pair of nodes in the WSNs, calculate the frequency of each node's appearance on all the shortest paths, and identify the segmentation nodes based on this frequency. The relevant parameters and their definitions are detailed in Table \uppercase\expandafter{\romannumeral1}.

 \begin{table}
 	\caption{Parameter List}
 	\label{table}
 	\setlength{\tabcolsep}{5pt}
 	\begin{tabular}{|p{50pt}|p{180pt}|}
 		\hline
 		Parameter & Definition\\
 		\hline
 		$N_i, N_j, N_k$& The nodes with identifiers $i$, $j$, and $k$\\
 		$P_{N_j,N_k}^g$& The set of nodes on the $g^{th}$ communication path between nodes $j$ and $k$\\
 		$SP_{N_j,N_k}^g$& The set of nodes on the shortest $g^{th}$ communication path between nodes $j$ and $k$\\
 		$|P_{N_j,N_k}^g|$& The number of hops on the $g^{th}$ path between nodes $j$ and $k$\\
 		$|SP_{N_j,N_k}^g|$& The number of hops on the shortest $g^{th}$ path between nodes $j$ and $k$\\
 		$TS_{i}$& The number that node $i$ appears on the shortest path between any two nodes\\
 		$n$& The total number of nodes in the localization scenario\\
 		$C_k$& The cluster $k$ including a group of nodes\\
 		$\mu_k$& The center point of the cluster $k$\\
 		$PN_i$& The node pair $i$ composed of two segmentation nodes\\
 		$N_a,N_b$& The segmentation nodes with node identifiers $a$ and $b$\\
 		$||N_a-N_b||$& The distance between segmentation node $a$ and segmentation node $b$\\
 		$L$ & The maximum distance that two nodes can communicate in the localization scenario\\
 		$U$& The bisector created based on segmentation nodes $N_a$ and $N_b$ in each pair\\
 		$HOP_{N_i,N_a}$& The number of hops from node $i$ to segmentation node $a$\\
 		$A_s, A_{w+1}$& The set of nodes in the area with identifiers $s$ and $w+1$\\
 		$OH^{A_s}_{N_i}$ & The number of one-hop nodes of node $i$ in the area $s$\\
 		$N_o, N_x, N_y$ & The reference nodes with node identifiers $o$, $x$, and $y$\\
 		$d_{ox}$ & The distance between node $o$ and node $x$\\
 		$\mathbb{R}$ & The Cartesian reference coordinate system within the area\\
 		$KN_u$& The node $u$ whose coordinates are known in the reference coordinate system\\
 		$x_u^{\mathbb{R}},y_u^{\mathbb{R}}$& The x-axis and y-axis coordinates of node $u$ whose coordinates are known in the reference coordinate system\\
 		$\mathbb{T}$ & The Cartesian globe coordinate system within the area\\
 		$x_u^{\mathbb{T}},y_u^{\mathbb{T}}$& The x-axis and y-axis coordinates of node $u$ whose coordinates are known in the globe coordinate system\\
 		$\Delta x,\Delta y$& The translation coefficients between the origin of the relative coordinate system and the origin of the global coordinate system\\
 		$R_1, R_2, R_3, R_4$& The rotation coefficients used to describe the orientation of the relative coordinate system with respect to the global coordinate system\\
 		\hline
 	\end{tabular}
 	\label{tab1}
 \end{table}

\subsection{Determining the shortest path between each pair of nodes}

\begin{figure}[htbp]
	\centerline{\includegraphics[width=3in]{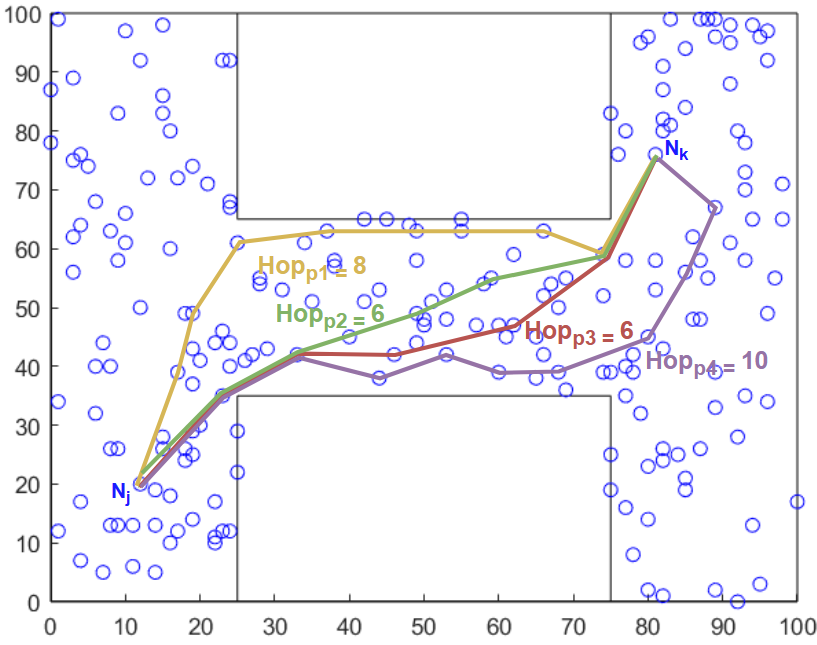}}
	\caption{Multiple communication paths can be established between each pair of nodes. Taking nodes $j$ and $k$ as an example, their communication paths are depicted in yellow, green, red, and violet. Among all these paths, the ones with the fewest hops are selected, as represented in green and red.}
	\label{f4}
\end{figure}

Due to the limited wireless radio transmission and energy save purpose, the data packets will be disseminated from a sender to a receiver in a multi-hop transmission pattern throughout the network \cite{6429030}. The data packets transmitted via multi-hops encapsulates the addresses of all relay nodes engaged in packet forwarding. Obviously, there will be a set of intermediate nodes on the communication path between every pair of nodes in the network. If $N_j$ and $N_k$ are two arbitrary nodes $j$ and $k$ in the localization area where $j \neq k$, the paths established between them can be denoted as $P_{N_j,N_k}^{g}$, where $g\in [1, max]$. This tells that not only one but multiple communication paths between nodes $N_j$ and $N_k$ can be set up. This is because the routing established in WSNs are data-centric rather than IP-centric. Also because data packets are transmitted using broadcasting or flooding pattern, many redundant paths will be formed between the source node and the destination node. By defining the initial value of $SP_{Nj, Nk}^g$ to $SP_{Nj, Nk}^1$ where $SP_{Nj, Nk}^1$ = $P_{Nj, Nk}^1$,  the set of nodes on the shortest communication path between nodes $N_j$ and $N_k$ where $g$ increases from 2 can be expressed as:

\begin{equation}
SP_{N_j,N_k}^{g}=
\begin{cases}
P_{N_j,N_k}^{g} & |P_{N_j,N_k}^{g}|<|SP_{N_j,N_k}^{g-1}|\\
SP_{N_j,N_k}^{g-1} & |P_{N_j,N_k}^{g}|>|SP_{N_j,N_k}^{g-1}|\\
P_{N_j,N_k}^{g} \cup SP_{N_j,N_k}^{g-1} & |P_{N_j,N_k}^{g}|=|SP_{N_j,N_k}^{g-1}|\\
\end{cases}
\label{7}
\end{equation}

The symbol $| \; |$ in the judgment criteria of Equation \ref{7} represents the number of hops on a path between nodes $j$ and $k$. Upon parameter $g$ reaches the maximum value $max$,  this iterative equation signifies that all the communication paths between nodes $j$ and $k$ have been exhaustively enumerated. After the iterative calculation, the shortest communication path between nodes $N_j$ and $N_k$ is found to be $SP_{N_j,N_k}^{max}$. Similarly, the shortest communication path between every pair of nodes in the network can be determined.

It is worth mentioning that the number of hops on different paths between a pair of nodes may be identical, which means there may be multiple shortest paths existing between each pair of nodes. This situation resembles the third scenario described in Equation \ref{7} for the piecewise function, where relay nodes along distinct paths with the same hop count will be comprehensively recorded. An example illustrating the various paths between nodes $j$ and $k$, out of which some of these paths share the same number of hops is given in Figure \ref{f4}.

\subsection{Calculating the number of occurrences of each node on the shortest path}

To identify the segmentation nodes which will be used for the network partitioning, the number of each relay node that appears on the shortest path between each pair of nodes in the network must be calculated. The number of node $i$, $N_i$,  occurring on the shortest communication path between each two nodes in the WSNs, $TS_i$, can be expressed as:

\begin{equation}
TS_{i} = \sum_{j=1}^{n-1} \sum_{k=j+1}^{n}
\begin{cases}
1 & N_i \in SP_{N_j,N_k}^{max}\\
0 & N_i \notin SP_{N_j,N_k}^{max}\\
\end{cases}
\label{eq}\end{equation}

\begin{figure}[htbp]
\centerline{\includegraphics[width=3.5in]{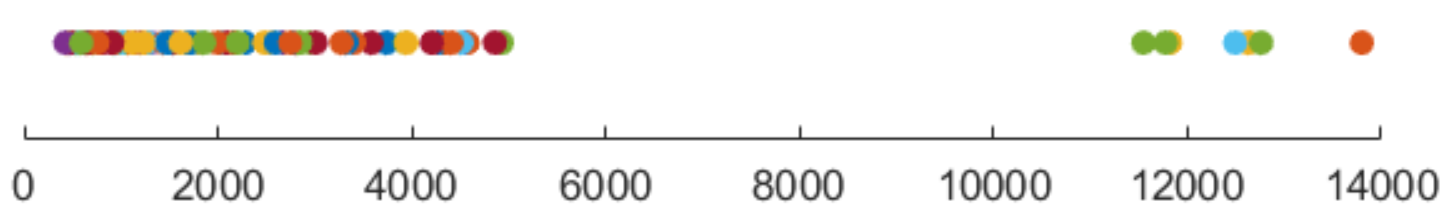}}
\caption{ The number of occurrence of each relay node on the shortest communication path between every pair of nodes in the network with the presence of H-shaped obstacle}
\label{f5}
\end{figure}

As shown in Figure \ref{f5}, the various colored points on the numerical axis indicate the frequency of occurrence for each relay node on the shortest paths between pairs of nodes. The H-shaped obstacle in the localization scenario influences these paths, resulting in a higher frequency of certain nodes being traversed. Consequently, a small cluster of points appears at the right end of the numerical axis, while the majority of points are concentrated at the left end, highlighting a significant numerical disparity between these two clusters. 

\begin{figure}[htbp]
	\centerline{\includegraphics[width=3.5in]{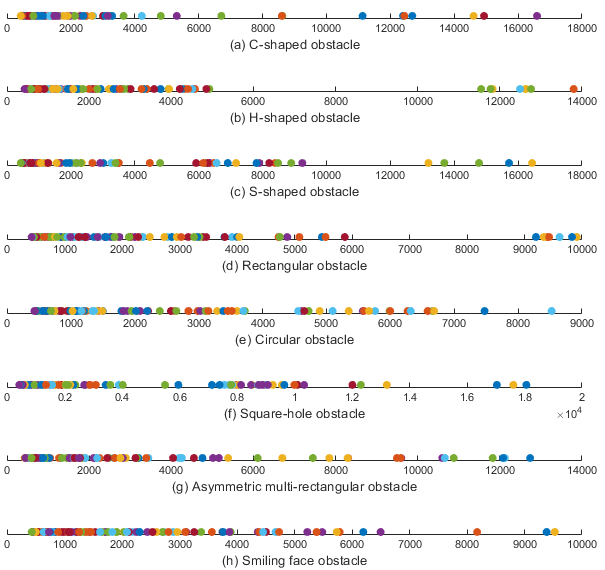}}
	\caption{ The number of occurrence of each relay node on the shortest communication path between every pair of nodes in the network with the presence of C-shaped, H-shaped, S-shaped, rectangular, circular, square-hole, asymmetric multi-rectangular, and smiling face obstacles}
	\label{f6}	
\end{figure}

However, this disparity is observed specifically in the analysis of the H-shaped obstacle. In fact, the distribution of the colored points may exhibit an intermittent pattern, influenced by the number of nodes located on the left and right, and the up and down sides of the convex corner of the obstacle, as well as the smoothness of that corner. Variations in obstacle shapes will result in varying number of occurrence of each relay node on the shortest path between every pair of nodes in the network, which may cause the boundaries between the two clusters to become blurred. This phenomenon is described in Figure \ref{f6} where the shapes of the obstacle is C, H, S, rectangle, circle, square-hole, asymmetric multi-rectangle, and smiling face, respectively. The shapes of these different obstacles are shown in Figures \ref{f16}-\ref{f23} in the section of experimental result.

Nevertheless, our goal is only to select a limited number of segmentation nodes, which will be definitely shown as the colored points at the right end of the numerical axis. This means that whether or not a numerical disparity shows up, and regardless of the size of that disparity, it will not impact the determination of the segmentation nodes.

\subsection{Identifying segmentation nodes}

Since obstacles are undetectable by sensor nodes in real-world WSNs, factors regarding the obstacles, such as their shape, position, quantity, and deployment patterns are unpredictable. Without these specifics, pre-defining the exact number of segmentation nodes becomes challenging. However, as only a limited number of the segmentation nodes need to be selected, it is  enough to learn which cluster contains the nodes with the largest number of occurrences on the shortest communication path between each pair of nodes in the network. To this end, the K-means \cite{electronics9081295} clustering algorithm will be employed to categorize nodes into two partitions. This clustering will help identify the nodes most frequently traverse the obstacle on the shortest paths, optimizing segmentation node selection based on frequency.

To start with, the occurrence counts for two arbitrary nodes on the shortest paths between each pair of nodes are selected as the initial center points for cluster 1, $C_1$ and cluster 2, $C_2$, where these center points are denoted as $\mu_1$ and $\mu_2$. For each node $N_i$ where $i \in [1, n]$, it will determine the cluster it belongs to using formula \ref{e9}.

\begin{equation}
\label{e9}
N_{i} \in
\begin{cases}
C_1 & |TS_i - \mu_1| \leq |TS_i - \mu_2|\\
C_2 & |TS_i - \mu_1| > |TS_i - \mu_2|\\
\end{cases}
\end{equation}

\vspace{0.08in}

\begin{equation}
	\mu_k = \frac{1}{|C_k|}\sum_{N_{1}^k}^{N_{m_k}^{k}} TS_i
\label{e10}
\end{equation}

\begin{figure}[htbp]
	\centerline{\includegraphics[width=3.5in]{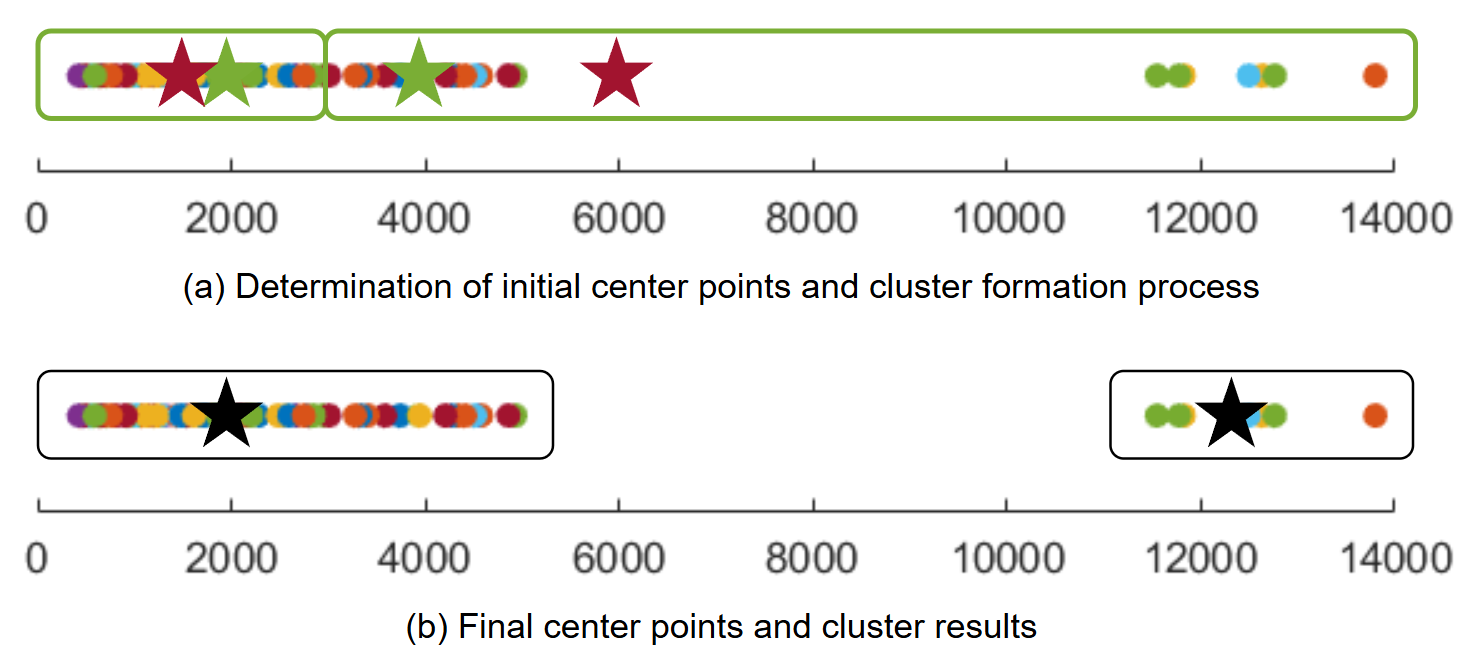}}
	\caption{The two green stars represent the initially randomly selected cluster center points.  The two red stars represent the new cluster center points calculated from the two clusters enclosed by the green rectangles.}
	\label{f7}
\end{figure}

After $C_1$ and $C_2$ are initially formed, each consisting of a group of nodes, the center points $\mu_k$ will be updated according to the relationship between the total number of nodes included in $C_k$, denoted as $|C_k|$, and the summation of occurrence counts of all nodes $N_i$ within $C_k$, where $k \in [1, 2]$, $i \in [1, m_k]$, and $m_k$ is the last node in $C_k$, as expressed by Equation \ref{e10}. Term $max^\mu$ is set to 1 if $(\mu_1\geq \mu_2)$ and to 2 otherwise.

Due to the iteration of the value of $\mu_k$ in Equation \ref{e10}, the coverage of clusters $C_1$ and $C_2$ will be updated accordingly. The iteration in Equations \ref{e9} and \ref{e10} will terminate only when the value of $\mu_k$ does not change any more.

An example illustrating the determination of center points and the formation of clusters is shown in Figure \ref{f7}(a). The two initial center points are represented by green stars, and the clusters formed around these points are enclosed within green rectangles. The red stars represent the updated center points, which are calculated based on the nodes in each of the two clusters. As the iteration runs, when the value of $\mu_k$ reaches stability, the center points of clusters are finalized and the cluster coverage ranges are fixed, as shown in Figure \ref{f7}(b).

\begin{figure}[htbp]
	\centerline{\includegraphics[width=3in]{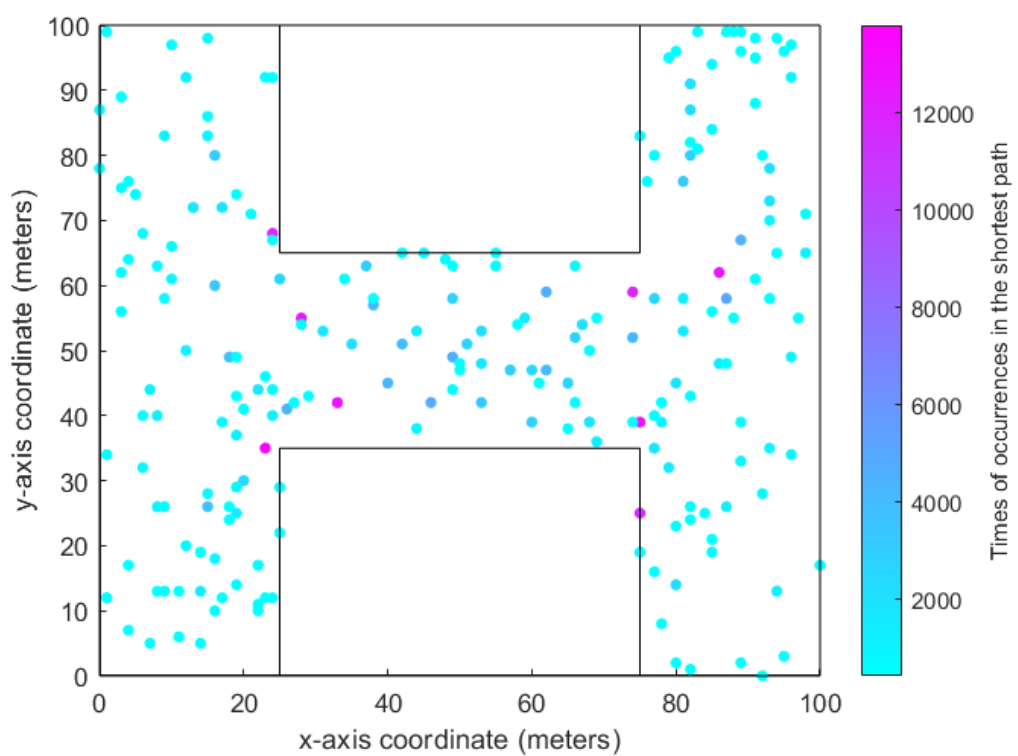}}
	\caption{The color map on the right represents the frequency with which each node appears on the shortest path between any two nodes. After applying the K-means clustering algorithm, eight purple nodes are ultimately identified as the segmentation nodes.}
	\label{f8}
\end{figure}

Finally, the nodes in the cluster with higher value of center point will be considered as the segmentation nodes. The distribution of the segmentation nodes in the presence of a H-shaped obstacle is shown in Figure \ref{f8}. It can be observed that all the discovered segmentation nodes are located near all the convex corners of the H-shaped obstacle.

Under normal circumstances, the number of nodes in the cluster with higher center point will generally be greater than or equal to the required number of segmentation nodes for a convex corner of the obstacle, which has been confirmed by many simulations. However, when the convex corner of the obstacle is located near the network boundary, it is possible that the number of nodes in the right cluster may be smaller than the required number of segmentation nodes. In such special cases, the identification of segmentation nodes can be disregarded. This is because partitioning the network based on such nodes, even if they can be found, would result in sub-networks, in which there will not exist sufficient nodes.  If there are not enough nodes in the sub-network, it becomes impossible to calculate the positions of unknown nodes, as typically at least three non-collinear anchor nodes are needed in the same area.

\section{Network Partitioning}

In this section, once the segmentation nodes are identified, which are typically located near each convex corner of the obstacle, they will be grouped into pairs. On this basis, the bisector of the segmentation nodes in each pair at the exterior angle of each convex corner of the obstacle will be evaluated. To calculate each bisector, all the nodes in the network that have the same distance to the segmentation nodes in each pair will be considered as part of the bisector for that specific convex corner of the obstacle. After building these bisectors for all the convex corners of the obstacle, the network can be partitioned into sub-networks, each free of any convex corners of the obstacle. This ensures that each sub-network avoids interference from the obstacle’s convex geometry, guaranteeing uninterrupted communication paths between nodes and facilitating efficient network organization for unknown node localization.

\subsection{Establishing pairs of segmentation nodes}

\begin{figure}[htbp]
\centerline{\includegraphics[width=3in]{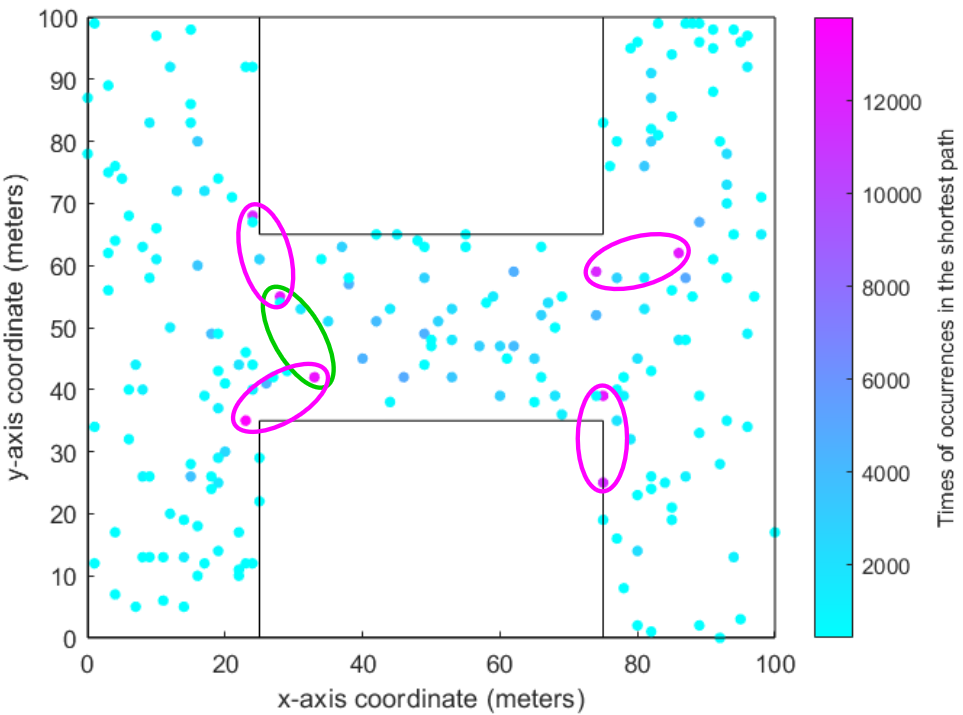}}
\caption{In the process of establishing segmentation node pairs, five node pairs meet the one-hop distance requirement. These initial five node pairs are highlighted by circles. Due to the fact that the two segmentation nodes within the green circle are simultaneously contained within two different adjacent pink circles, these two nodes within the green circle will be excluded from being used for network partitioning. }
\label{f9}
\end{figure}

Segmentation nodes that can communicate directly with each other will form a pair, denoted as $PN_i$. A pair of segmentation nodes should satisfy the condition that the two segmentation nodes are within one-hop of each other, meaning the distance between nodes $N_a$ and $N_b$ must be less than or equal to the maximum radio transmission range, $||N_a - N_b|| \leq L$, where $N_a, N_b \in C_{max^\mu}$, $N_a \neq N_b$, and $L$ represents the one-hop distance. The initial formation of segmentation node pairs for all the convex corners of the H-shpaed obstacle is described in Figure \ref{f9}. 

Since the network will be divided into parts according to the bisectors, which will be established based on the segmentation node pairs, the bisectors established based on the overlapped segmentation node pairs will cause excessive division of the network. If the network is divided into unnecessary several sub-networks, the number of nodes in some sub networks may not be sufficient and there may appear convex corners of the obstacle in some of the sub-networks.

As can be seen in the figure, it is possible for a segmentation node to be included in several pairs, resulting in overlapping regions between circles representing different segmentation node pairs, as illustrated by the pink and green circles. Since the network will be partitioned based on the bisectors which are created from these segmentation node pairs, the bisectors formed from overlapping pairs could lead to excessive division of the network. Over-dividing the network into more sub-networks than necessary may result in some sub-networks lacking sufficient nodes. This can complicate effective partitioning, as certain sub-networks may not have enough nodes to evaluate their locations.

Therefore, if the segmentation nodes in a particular pair, such as the green circle in the figure, are also part of other pairs, and these other pairs to which these two segmentation nodes belong are different, that particular pair should be removed. To achieve this, for a segmentation node pair $PN_i$ consisting of nodes $N_a$ and $N_b$, if $N_a \in PN_j$ and $N_b \in PN_k$, where $i \neq j \neq k$, the pair $PN_i$ will be removed from the network. This prevents redundant divisions in the network.

It worth mentioning that if one of the segmentation nodes in a pair is also the node in another pair, neither pair will be removed. This avoids excessive deletion of valid segmentation node pairs that are still needed for network partitioning.

\subsection{Bisector establishment to transform obstacle from convex to concave}

Since the exterior angle of each convex corner of the obstacle ranges from 180° to 360°, nodes near these corners may experience poor communication quality due to the interference from the obstacle. To address this issue, a bisector will be established for each pair of segmentation nodes to partition the network into multiple regions. After the network is partitioned based on these bisectors, all the convex corners of the obstacle will be transformed into concave ones, where the exterior angle of a concave corner, from the perspective of nodes, ranges from 0° to 180°. In this way, the communication between nodes in each newly formed sub-network will no longer be obstructed by the obstacle.

\begin{figure}[htbp]
	\centerline{\includegraphics[width=2in]{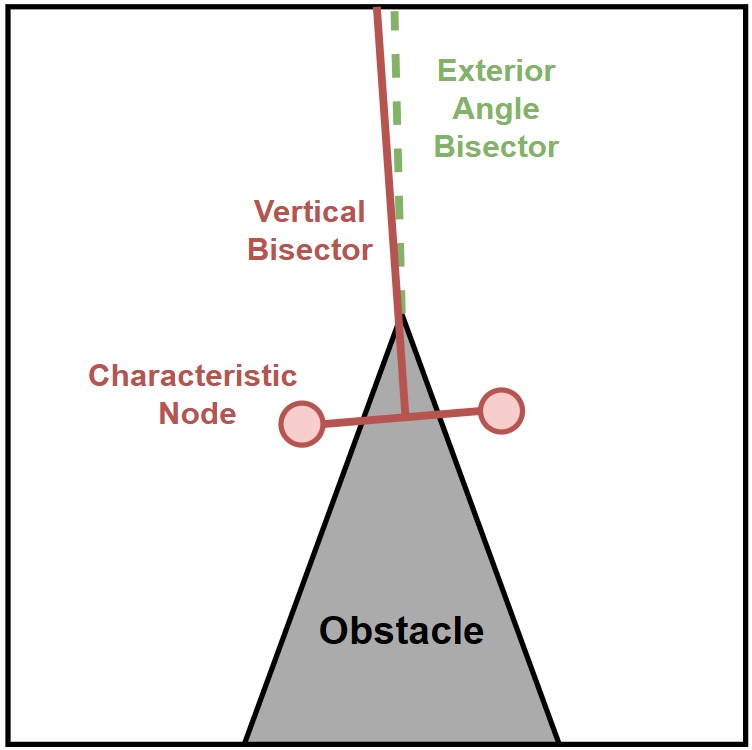}}
	\caption{The convex corner of the obstacle is transformed into a concave corner by creating bisectors, which are constructed based on the connection lines between each pair of segmentation nodes.}
	\label{f10}
\end{figure}

To establish bisectors, given that the shape, position, size, quantity, and layout of obstacles is hidden from sensor nodes, precisely halving the exterior angle of the obstacle's convex corner is impractical. Therefore, the bisector will be formed perpendicular to the line that connects the two segmentation nodes in each pair. The difference between this bisector and a theoretically ideal bisector that directly divides the convex corner’s exterior angle is small, as shown by the solid red line and the dash green line in Figure \ref{f10}.

Please note that the bisector line is not an actual line but rather a sequence of nodes that are equidistant from both segmentation nodes in the same pair. Together, these nodes form the bisectors for all the convex corners of the obstacle. For the two segmentation nodes in each pair, $N_a$ and $N_b$, the bisector will consist of all nodes $N_i$, in the region where $N_a$ and $N_b$ are located, that are equidistant from $N_a$ and $N_b$, meaning they have the same number of hops to both segmentation nodes. Thus, the bisector created based on segmentation nodes $N_a$ and $N_b$ in each pair can be defined as set $U$: 

\begin{equation}
U=\{N_i| \hspace{0.1cm} HOP_{N_i, N_a}=HOP_{N_i, N_b}\}
	\label{e11}
\end{equation}

In Equation \ref{e11}, terms $HOP_{N_i, N_a}$ and $HOP_{N_i, N_b}$ represent the number of hops from node $N_i$ to nodes $N_a$ and $N_b$, respectively. It is important to note that since network partitioning based on pairs of segmentation nodes is performed sequentially, the network may has been already partially partitioned before the partitioning process for a specific pair of segmentation nodes begins.

\subsection{Specifying the area where each node is located}

Although all segmentation nodes of convex corners of the obstacle are identified simultaneously,  the bisectors are established at different times due to the varying lengths of them. This is because the time required to find all equidistant nodes for each pair of segmentation nodes may differ. Consequently, the partitioning of each node in the network to a specific sub-network occurs at different times, meaning that the network partitioning process is conducted sequentially.

Only when the segmentation nodes in a pair belong to the same enclosed area, denoted as $A_s$, will all nodes within this area, $N_i$, be divided into two groups based on the bisector, with each group forming a distinct sub-network. Here, parameter $s$ serves as the identifier for the current sub-network, where $s \in [1, z]$, $z$ represents the last partitioned sub-network, and $A_1$ denotes the entire network, which is an enclosed area containing all the nodes. Parameter $w$ denotes the number of partitioning operations performed across the entire network up to now. Thus, for all the nodes in the same area $A_s$, the sub-networks to which they will be partitioned can be determined as:

\begin{equation}
	N_i \in
	\begin{cases}
		A_s & HOP_{N_i,N_a} < HOP_{N_i,N_b}\\
		A_{w+1} & HOP_{N_i,N_a} > HOP_{N_i,N_b}\\
	\end{cases}
	\label{e12}
\end{equation}

Please note that a pair of segmentation nodes also participate in the network partitioning organized by another pair of segmentation nodes. In other words, it may happen that before the network is partitioned based on a specific pair of segmentation nodes, these nodes have already been assigned to the two different sub-networks, which are the partitioning result from another pair of segmentation nodes. Therefore, in such cases, if the segmentation nodes are initially placed in different sub-networks, they will not arrange further network partitioning. This can prevent redundant partitioning when the bisector line for one pair of segmentation nodes intersects with the connection line of the segmentation nodes in another pair.

\begin{figure}[htbp]
	\centerline{\includegraphics[width=2in]{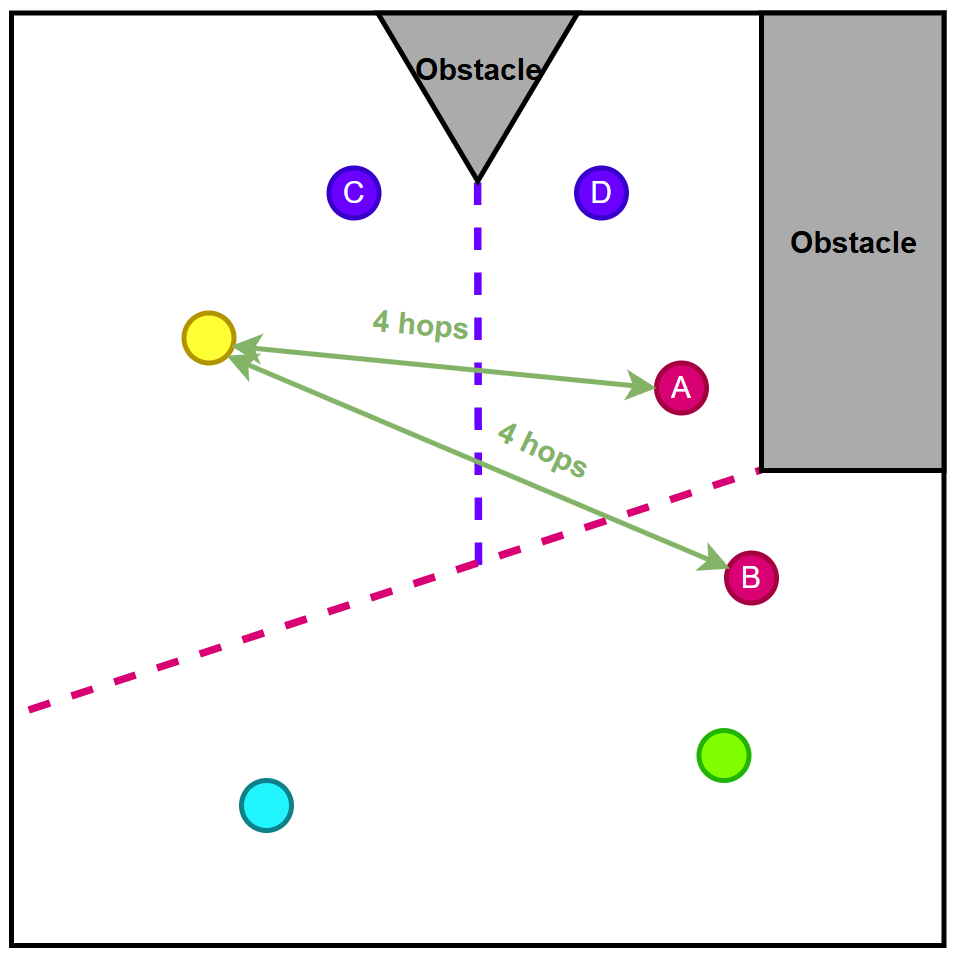}}
	\caption{Regardless of the order in which the bisectors are generated, the network partitioning results remain unaffected. This demonstrates that our designed algorithm is capable of automatically updating the sub-network to which nodes should belong.
	}
	\label{fig10+}
\end{figure}

The bisectors originating from different segmentation nodes may or may not intersect. When a bisector intersects with another, the search for equidistant nodes for this specific segmentation node pair will terminate, in order to avoid redundant network partitioning. Although the bisectors are formed sequentially and the network partitioning proceeds accordingly, the assignment of nodes to their destined sub-networks remains unaffected by these factors. This can be illustrated in Figure \ref{fig10+}, where the red and purple nodes, labeled with letters, represent two distinct segmentation node pairs. The red and purple dotted lines depict the approximate bisectors generated from the respective segmentation node pairs, with the red bisector formed before the purple one. Consequently, the network partitioning for the nodes will be first determined based on the red bisector and subsequently refined by the purple bisector. There may exist some nodes that are equidistant to both segmentation node pairs. For example, the yellow node is initially assigned to the upper sub-network according to the red bisector, and then reassigned to the upper-left sub-network based on the purple bisector. It can be observed that even if the purple sector is formed earlier than the red one, the final sub-network assignment of the yellow node will remain unchanged in essence. This partitioning principle also applies to the green and blue nodes, with the blue node sharing a sub-network with either the yellow or green node. This proves that the results of network partitioning are independent of the order in which the bisectors are generated and the sequence of network partitioning. As a result, the algorithm is capable of automatically updating the sub-network to which nodes belong, and thus does not affect the node localization accuracy.

\begin{equation}
	N_i \in
	\begin{cases}
		A_s & OH^{A_s}_{N_i} \geq OH^{A_{w+1}}_{N_i}\\
		A_{w+1} & OH^{A_{s}}_{N_i} < OH^{A_{w+1}}_{N_i}\\
	\end{cases}
	\label{e13}
\end{equation}

\begin{figure}[htbp]
	\centerline{\includegraphics[width=3.5in]{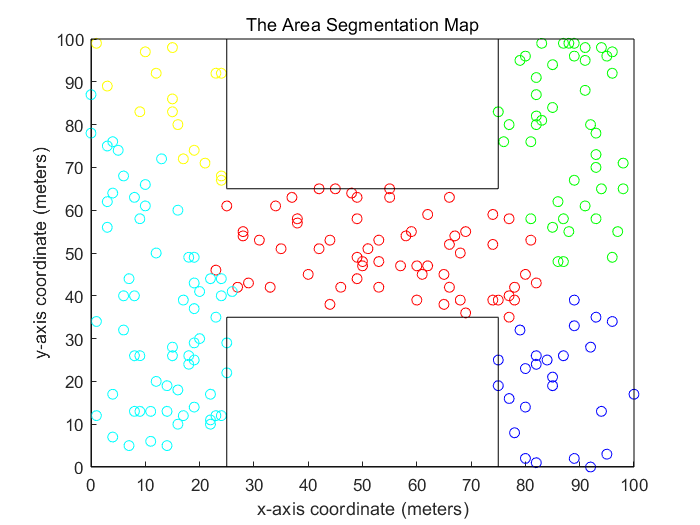}}
	\caption{Each node performs self-organizing partitioning based on the bisectors of segmentation nodes. Nodes that reside in distinct sub-networks are depicted with different colors, which represent the results of the network partitioning.}
	\label{f11}
\end{figure}

In Equation \ref{e12}, in addition to the normal nodes being divided into two sub-networks, the segmentation nodes $N_a$ and $N_b$ are also assigned to distinct sub-networks $A_s$ and $A_{w+1}$, where $N_a \in A_s$ and $N_b \in A_{w+1}$. After the nodes that are not on the bisectors have been partitioned, each node will learn the sub-network it belongs, and record the corresponding sub-network identifier $s$.

For all the nodes that are part of the bisector for a given segmentation node pair, $N_i \in U$, their assignment to a specific sub-network will depend on the number of one-hop neighbors they have in each of the two adjacent sub-networks, $OH^{A_s}_{N_i}$ and $OH^{A_{w+1}}_{N_i}$. Therefore, for such a node, $N_i$, the sub-network it will belong to can be determined using Equation \ref{f11}.

Finally, after the network is divided based on all the bisectors of the existing segmentation nodes, the network will be partitioned into $z$ sub-networks, ensuring no unnecessary partitioning occurs, whether fewer or more divisions than needed. The result of network partitioning with the presence of H-shaped obstacle is shown in Figure \ref{f11}.

\begin{algorithm}
	\caption{Determination of segmentation nodes and network partitioning}
	\begin{algorithmic}
		\STATE \bf For All Nodes:
		\STATE \rm Initialize node identifier $i$
		\STATE Determining the shortest path between each pair of nodes, denoted as $SP_{N_j,N_k}^{max}$
		\STATE $TS_i \gets 0$
		\FOR{$j \gets 1$ \bf{to} $n-1$}
		\FOR{$k \gets j+1$ \bf{to} $n$}
		\IF{$N_i \in SP_{N_j,N_k}^{max}$}
		\STATE $TS_i \gets TS_i + 1$
		\ENDIF
		\ENDFOR
		\ENDFOR
		\STATE \rm Broadcast $TS_i$ to each node
		\STATE Randomly select the $TS$ of two nodes as $\mu_1$ and $\mu_2$
		\REPEAT
		\IF {$|TS_i - \mu_1| \leq |TS_i - \mu_2|$}
		\STATE $N_i \in C_1$
		\ELSE
		\STATE $N_i \in C_2$
		\ENDIF
		\STATE Update $\mu_1$ and $\mu_2$ according to equation (10)
		\UNTIL {$\mu_1$ and $\mu_2$ remain constant}
		\IF {$\mu_1 \geq \mu_2$}
		\STATE $SN \gets C_1$
		\ELSE
		\STATE $SN \gets C_2$
		\ENDIF
		\IF {$N_i \in SN$}
		\FOR{$j \gets 1$ \bf{to} $n$}
		\IF{$i \neq j$ and $||N_i - N_j|| \leq L$}
		\STATE \rm $N_i$ and $N_j$ constitute segmentation node pairs
		\ENDIF
		\ENDFOR
		\ENDIF
		\IF {$N_i$ and $N_j$ are respectively in other segmentation node pairs}
		\STATE Cancel the segmentation node pair of $N_i$ and $N_j$
		\ENDIF
		\REPEAT
		\IF {A segmentation node pair consisting of $N_a$ and $N_b$ is located within the same area}
		\IF {$HOP_{N_i,N_a} < HOP_{N_i,N_b}$}
		\STATE $N_i \in A_s$
		\ELSIF{$HOP_{N_i,N_a} > HOP_{N_i,N_b}$}
		\STATE $N_i \in A_{w+1}$
		\ELSE
		\IF{Surrounding node partitioning has been completed}
		\IF{$OH^{A_s}_{N_i} \geq OH^{A_{w+1}}_{N_i}$}
		\STATE $N_i \in A_s$
		\ELSE
		\STATE $N_i \in A_{w+1}$
		\ENDIF
		\ENDIF
		\ENDIF
		\ENDIF
		\UNTIL{All segmentation pairs have been utilized}
	\end{algorithmic}
\end{algorithm}

\section{Localization of Unknown Nodes}

All the designs discussed in the previous sections focus solely on the relative relationships between nodes. To determine the physical coordinates of all unknown nodes, the information of the reference coordinate system is essential. Although it is assumed that each sub-network contains at least three anchor nodes, which implies that a global coordinate system has been established, the localization of unknown nodes cannot be performed on a global scale. This limitation arises because distance measurements between one-hop nodes may be inaccurate, due to the presence of convex corners of obstacles in certain areas of the network. Therefore, in this section, a relative coordinate system will be first established, and the relative coordinates of unknown nodes will be determined separately within each sub-network, under the assumption that the network density is not sparse and nodes are randomly distributed. Finally, the relative coordinate system will be calibrated to the globe coordinate system, enabling the accurate determination of the physical coordinates of all unknown nodes across the entire network.



\subsection{Establishing a unique relative coordinate system for each sub-network}

The presence of anchor nodes in the network signifies that a globe coordinate system has been established for the WSN. To locate unknown nodes, anchor nodes will be utilized. The calculation process generally begin by identifying the location of an unknown node that is close to these anchor nodes and then gradually extend toward the network boundaries, incrementally determining the positions of all nodes. However, the accuracy of the initial unknown node’s location estimate largely depends on the distance to the anchor nodes. Given that the number of anchor nodes is much smaller than that of unknown nodes, it is probable that these anchor nodes will be separated by more than one hop.

In such a case, significant errors may arise. This is because the RSSI technique is used to approximate distances between communicating nodes, but it cannot accurately measure distances over multiple hops. When an unknown node is two or more hops from the other two anchor nodes, its distance to these anchor nodes will be estimated based on hops, which typically results in an overestimation of the true distance. Consequently, this unknown node may only rely on the nearest anchor node within one hop for positioning, which is insufficient for precise localization, as the RSSI value alone does not provide directional information. Therefore, precise localization of this initial unknown node cannot be achieved, which ultimately affects the accuracy of the location estimates for all unknown nodes in the network.

Even if three anchor nodes are positioned within one hop of each other, network-wide localization of unknown nodes can still encounter inaccuracies, although the initial unknown node's position can be accurately determined due to its one-hop proximity to two of these anchor nodes. Obstacle with convex corners can further degrade signal quality, since this interference will disrupt the RSSI-based estimation of distances between nodes, leading to cumulative errors in the estimated relative positions of all nodes across the network. Thus, obstacles in the environment negatively affect the reliability of RSSI for node positioning.

Thus, the localization of unknown nodes can only be carried out within partitioned sub-networks, with the assumption that each sub-network contains at least three anchor nodes. However, since the anchor nodes within the same sub-network may be separated by more than one hop, distance estimates between an unknown node and each anchor node may still lack accuracy. As a result, the global coordinate system for the entire network will not be used to calculate unknown node locations. Instead, separate and independent relative coordinate system will be established for each sub-network, allowing for more accurate and localized positioning within each sub-network.

To this end, three nodes $N_o, N_x$, and $N_y$ will be selected within each sub-network, where these nodes can be anchor nodes, unknown nodes, or a mix of both, which should meet the following criteria:

\noindent(a) $HOP_{N_o,N_x} = HOP_{N_o,N_y} = HOP_{N_x,N_y} = 1$

\noindent(b) $d_{ox}+d_{xy}\textgreater d_{yo}$, $d_{ox}+d_{yo}\textgreater d_{xy}$, and  $d_{xy}+d_{yo}\textgreater d_{ox}$

\noindent(c) $min[\max_{1 \leq i \leq n^{s}}{(HOP_{N_o,N_i} \!+\!HOP_{N_x,N_i} \! + \! HOP_{N_y,N_i})}]$

For criterion (a), it is required that the number of hops between each pair of the three reference nodes be exactly one. This ensures that in each sub-network, the worst-case scenario is that there will be a node located one hop from two of the reference nodes and more than one hop away from the third, if the network density is not too sparse. In this way, the position of this particular unknown node can be calculated, which serves as the starting point for the localization process of unknown nodes within the sub-network.

For criterion (b), the three selected nodes must not be collinear, meaning the line connecting them cannot form a straight line. This ensures that the position of a node closest to these three reference nodes can be uniquely determined, avoiding ambiguity in localization.

For criterion (c), each sub-network must have multiple set of three nodes that satisfy both criteria (a) and (b). For each set of three nodes, there will be a node in the same sub-network $s$, denoted as $N_i$ where $i \in [1, n^s]$ and $n^s$ is the last node in sub-network $s$, that is farthest from these three nodes. This farthest node can be identified by calculating the sum of the hop counts between each of the three nodes and itself. If the summation yields the maximum value, the corresponding $N_i$ is selected as the farthest node to this set of three nodes. Similarly, the summation of the hop counts for all sets of three nodes and their respective farthest node in each sub-network can be calculated. Among all these summation values, only the three nodes which are associated with this minimum summation value will be determined as the reference nodes, which are used as the basis for unknown node localization within each specific sub-network.

Please note that there may not be only one set of nodes within the same area satisfy the selection criteria (a), (b), and (c). However, only one of them is chosen to be the reference nodes for that sub-network.

\begin{figure}[h]
	\centerline{\includegraphics[width=3.5in]{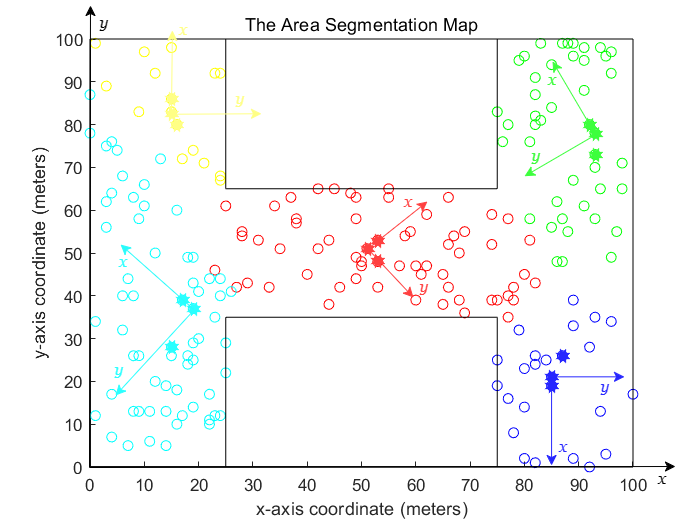}}
	\caption{The three reference nodes in each sub-network are depicted as eight-pointed stars, with distinct colors used to differentiate the reference nodes across various sub-networks. The Cartesian coordinate system established within each sub-network is illustrated using two vectors of matching colors, corresponding to the respective reference nodes in the same sub-network.}
	\label{f12}
\end{figure}

For each sub-network, a Cartesian coordinate system $\mathbb{R}$ is constructed based on the three reference nodes. Node $N_o$ serves as the origin of the coordinate system, with the half-line extending from $N_o$ to $N_x$ defined as the x-axis. The orientation of the y-axis is determined using $N_y$, ensuring orthogonality to the x-axis. The distance between $N_o$ and $N_x$, $||N_o - N_x||$ is employed to establish the unit length for both coordinate axes. By measuring the distance to node $N_o$ using the RSSI technique, the positions of the reference nodes $N_x$ and $N_y$ can be precisely determined. Figure \ref{f12} illustrates the reference nodes and the corresponding coordinate system defined within each sub-network.

\subsection{Locating unknown nodes in relative coordinate system of each sub-network}

The localization of all unknown nodes' coordinates in each sub-network begins by locating the unknown node, $N_i$, that is closest to the three reference nodes. These reference nodes constitute the known-node set, denoted as $KN_u$, where $u \in [1, max_r]$ and $max_r$ is initially set to 3, with coordinates $(x_u^{\mathbb{R}}, y_u^{\mathbb{R}})$. Depending on whether $N_i$ lies within the radio coverage of one, two, or all three reference nodes, which is related to the density and deployment of sensor nodes in the network, the coordinates of $N_i$ can be evaluated as:

\subsubsection{$N_i$ is within the radio range of three reference nodes}

In this case, the position of $N_i$ can be determined using trilateration. By treating the coordinates of the three reference nodes as the centers of circles and using the measured relative distances as their corresponding radii, three circles can be constructed. The unique location of $N_i$ will be at the intersection of these three circles. Hence, the coordinates $(x_i, y_i)$ of $N_i$ satisfy Equation \ref{e14}:
		
\begin{equation}
(x_i-x_u^{\mathbb{R}})^2 + (y_i-y_u^{\mathbb{R}})^2 = ||N_i-KN_u||^2
\label{e14}	
\end{equation}

\subsubsection{$N_i$ is within the radio range of two reference nodes}

There are two possible scenarios when $N_i$ falls within the communication range of two reference nodes. In the first scenario, the circles created by the two reference nodes are tangent to each other. This results in a single intersection point, indicating that the location of $N_i$ can still be accurately determined.

In the second scenario, the two circles have two intersection points, as they partially overlap. This requires assistance from the third reference node. Since $N_i$ is outside the radio range of the third reference node, the intersection point which is farther from the third reference node will be regarded as the coordinates of unknown node $N_i$.

\subsubsection{$N_i$ is within the radio range of one reference node}

This situation typically arises when the unknown node is positioned at the edges of a sub-network. In such cases, the unknown node $N_i$ should be positioned at some point on the boundary of the circle centered at the single reference node it can reach. Therefore, determining the precise location of $N_i$ needs help from at least one other reference node. The communication ranges of these reference nodes help exclude parts of the circle's arc that are inconsistent with $N_i$'s possible position. This effectively narrows the potential locations for $N_i$ to a smaller segment of the arc. The midpoint of this narrowed arc is then selected as the estimated coordinates of $N_i$ within the relative coordinate system $\mathbb{R}$.

The determination of the coordinates of unknown node $N_i$ in the three aforementioned cases is illustrated in Figure \ref{f13}. The localization of unknown nodes should first be performed when their positions can be uniquely determined, as in the first two cases. Once all such unknown nodes have been localized, the remaining unknown nodes in the third case will be determined based on the exclusive arc. Once the location of an unknown node 
is identified, it becomes a new member of the known node set $KN_u$, and the value of $max_r$ is incremented by one. Through this iterative process, the coordinates of all unknown nodes within each sub-network will be progressively determined in their respective coordinate system $\mathbb{R}$.

\begin{figure}[h]
	\centerline{\includegraphics[width=2.5in]{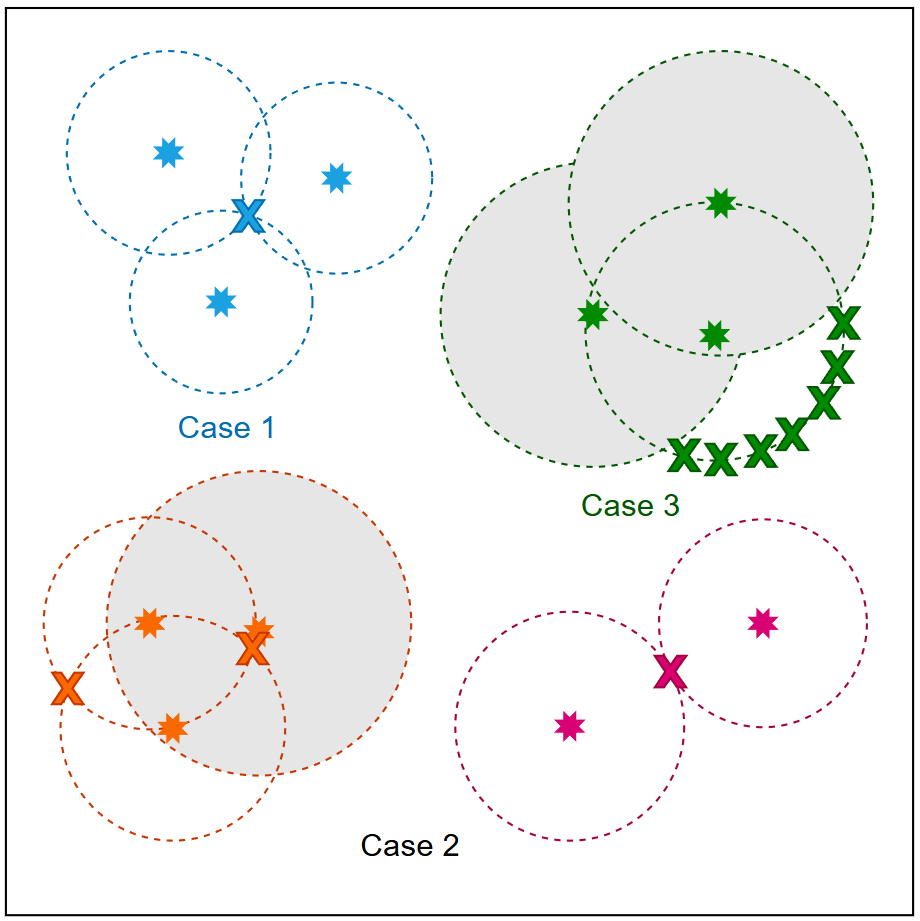}}
	\caption{The coordinates of an unknown node in each sub-network are determined based on its communication range with the reference nodes. For case 1, the unknown node is within the range of all three reference nodes, allowing precise localization. For case 2, the unknown node is within the range of two reference nodes. For the left case, the two circles are tangent, resulting in one intersection point. For the right case, the circles overlap, producing two intersection points, with the farther point selected as the node's location. For case 3, only one reference node is reachable. With the help of the other two reference nodes, the midpoint of the non-excluded arc can be calculated.}
	\label{f15}
\end{figure}

\subsection{Calibrating the relative coordinate system to globe coordinate system}

The calibration of the relative coordinate system to the globe coordinate system is performed independently for each sub-network, optimally utilizing anchor nodes that are accurately localized through the first two cases described in the previous section, under the assumption that each sub-network contains at least three non-collinear anchor nodes. Given the locations of anchor nodes in the globe coordinate system  $\mathbb{T}$ as $(x_u^\mathbb{T}, y_u^\mathbb{T})$, and in the relative coordinate system $\mathbb{R}$ as $(x_u^\mathbb{R}, y_u^\mathbb{R})$, where $u\in [1,max^a]$ and $max^a$ denotes the maximum number of anchor nodes in each sub-network, the translation coefficients $\Delta x$ and $\Delta y$ representing the offset between the origin of the relative coordinate system and the origin of the global coordinate system, as well as the rotation coefficients $R1$, $R2$, $R3$, and $R4$ describing the orientation of the relative coordinate system with respect to the global coordinate system, can be determined, as shown in Equation \ref{e15}. These coefficients enable all unknown nodes within each sub-network to localize themselves and resolve their absolute coordinates in the globe coordinate system.

\begin{equation}
	\left[
	\begin{matrix}
		x^{\mathbb{T}}_u \\
		y^{\mathbb{T}}_u \\
	\end{matrix}
	\right]
	=
	\left[
	\begin{matrix}
		R_1&R_2& \Delta x \\
		R_3&R_4&\Delta y \\
	\end{matrix}
	\right]
	\left[
	\begin{matrix}
		x^{\mathbb{R}}_u \\
		y^{\mathbb{R}}_u \\
		1 \\
	\end{matrix}
	\right]
	\label{e15}
	\end{equation}


%

\section{Numerical Results}

This section evaluates the localization accuracy of unknown nodes under eight different scenarios, each featuring obstacles with varying shapes, layouts, quantities, positions, and sizes, which are all unknown to sensor nodes. For each scenario, simulations are conducted using four configurations with different numbers and ratios of unknown nodes to anchor nodes. To assess the effectiveness of network partitioning, the number of convex corners in the obstacles, the theoretically ideal number of partitions, and the actual number of partitions are compared across the four node configurations. The partitioning performance is further evaluated by computing the ratio of node pairs that communicate by traversing obstacles after partitioning compared to before. Finally, localization accuracy is evaluated by analyzing the number of nodes with incorrect position estimates, along with the positional error between the actual and estimated locations of unknown nodes in each scenario.

\subsection{Simulation Environment Setups}
The proposed algorithm is simulated in MATLAB with version R2022b using the standard computational library. To systematically evaluate the partitioning accuracy and localization performance under diverse environmental conditions, the simulations are conducted in a defined $100 \times 100 m^2$ network area \cite{DESA2016322}. 

\begin{figure*}[htbp]  
	\centering
	\begin{minipage}[t]{0.49\textwidth}
		\centering
		\subfigure[]{\raisebox{-\height}{\includegraphics[scale=0.17]{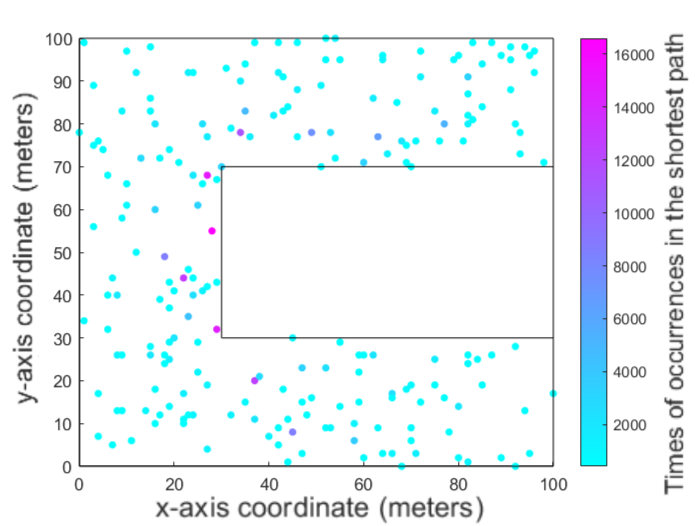}}}
		\subfigure[]{\raisebox{-\height}{\includegraphics[scale=0.17]{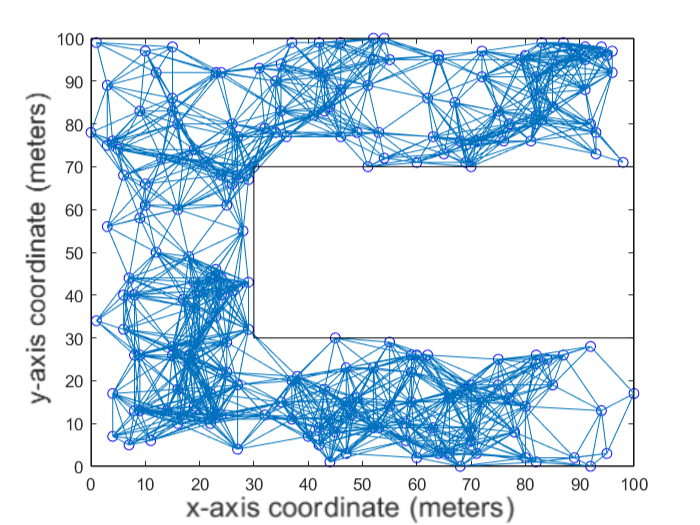}}}
		\subfigure[]{\raisebox{-\height}{\includegraphics[scale=0.17]{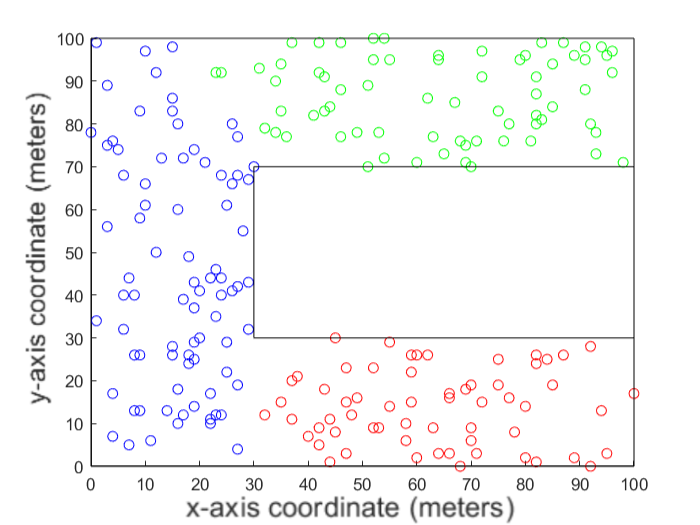}}}
		\subfigure[]{\raisebox{-\height}{\includegraphics[scale=0.17]{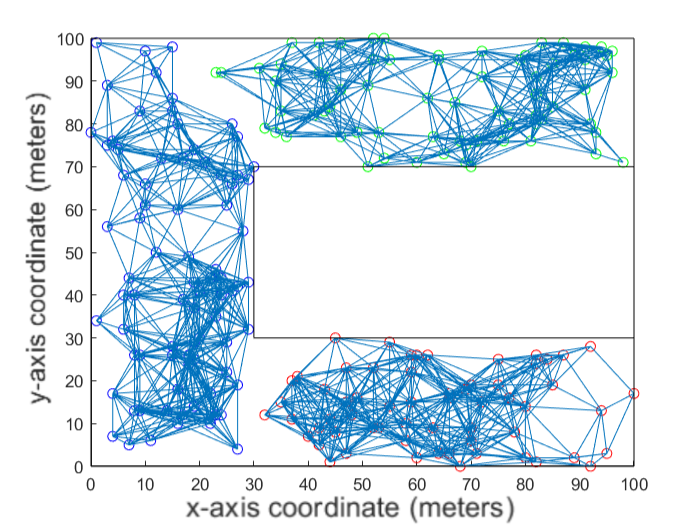}}}		
		\vspace{-0.2cm}
		\caption{C-shaped obstacle}
		\label{f16}
	\end{minipage}
	\hfill
	\begin{minipage}[t]{0.49\textwidth}
		\centering
		\subfigure[]{\raisebox{-\height}{\includegraphics[scale=0.17]{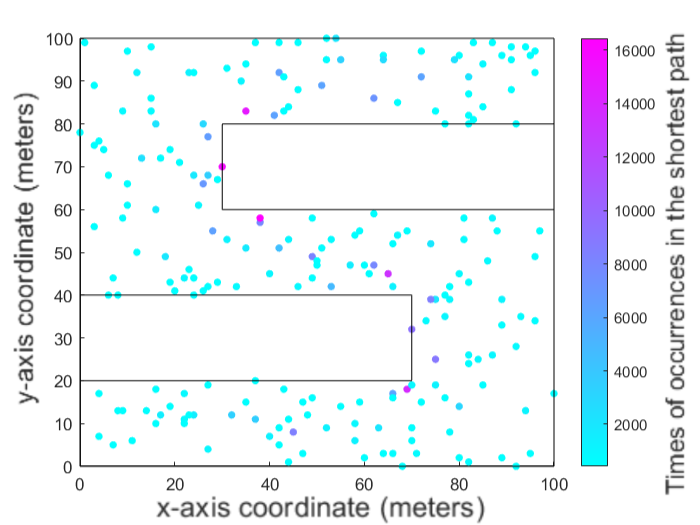}}}
		\subfigure[]{\raisebox{-\height}{\includegraphics[scale=0.17]{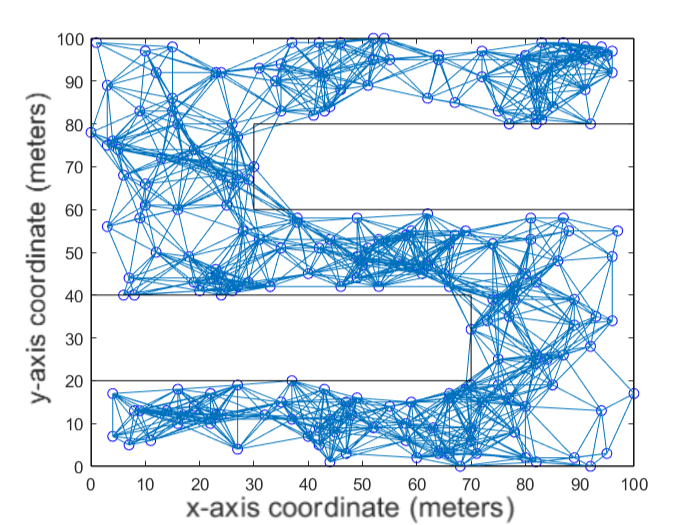}}}
		\subfigure[]{\raisebox{-\height}{\includegraphics[scale=0.17]{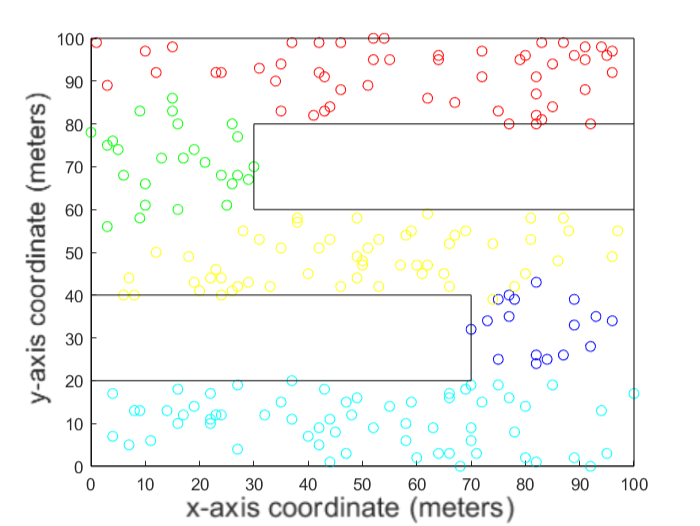}}}
		\subfigure[]{\raisebox{-\height}{\includegraphics[scale=0.17]{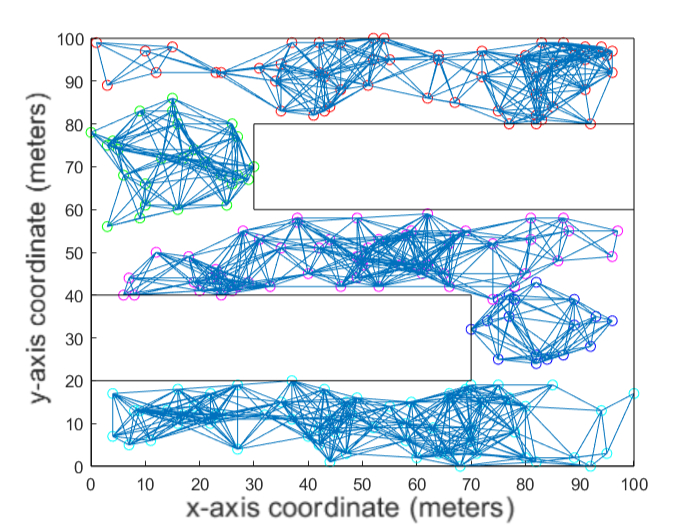}}}
		
		\vspace{-0.2cm}
		\caption{S-shaped obstacle}
		\label{f17}
	\end{minipage}
	
	\vspace{0.8cm}
	\begin{minipage}[t]{0.49\textwidth}
		\centering
		\subfigure[]{\raisebox{-\height}{\includegraphics[scale=0.17]{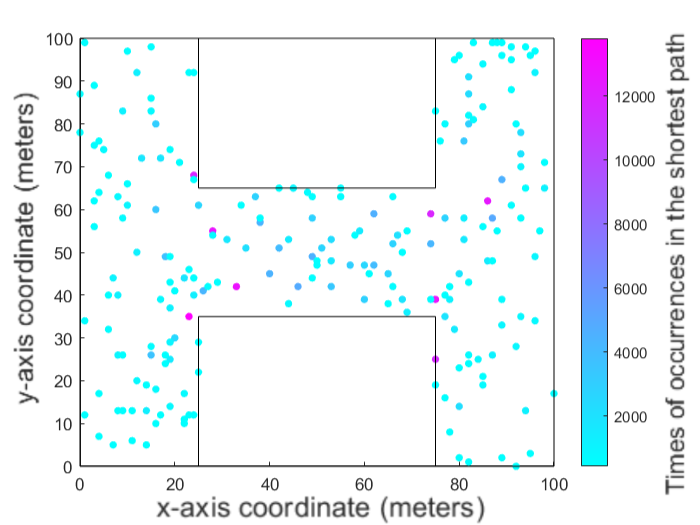}}}
		\subfigure[]{\raisebox{-\height}{\includegraphics[scale=0.17]{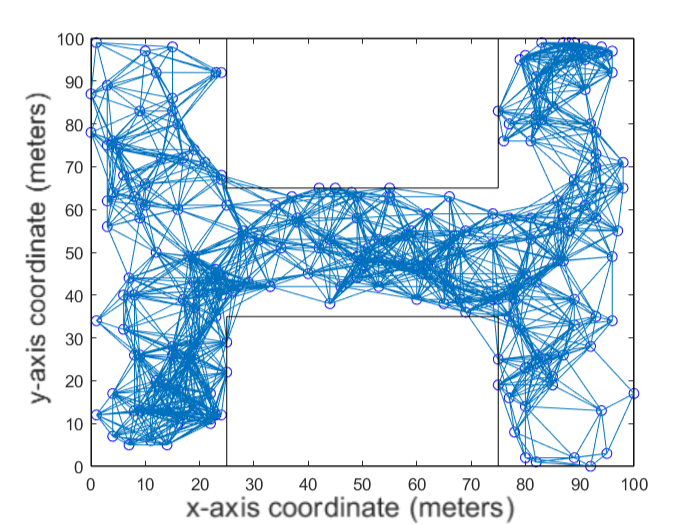}}}
		\subfigure[]{\raisebox{-\height}{\includegraphics[scale=0.17]{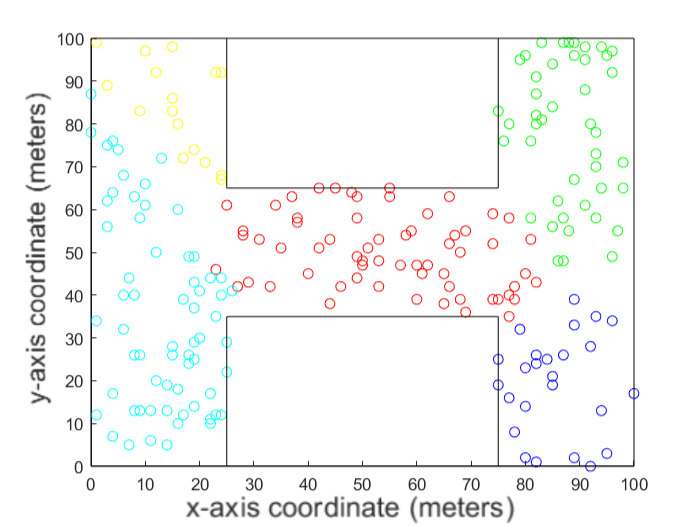}}}
		\subfigure[]{\raisebox{-\height}{\includegraphics[scale=0.17]{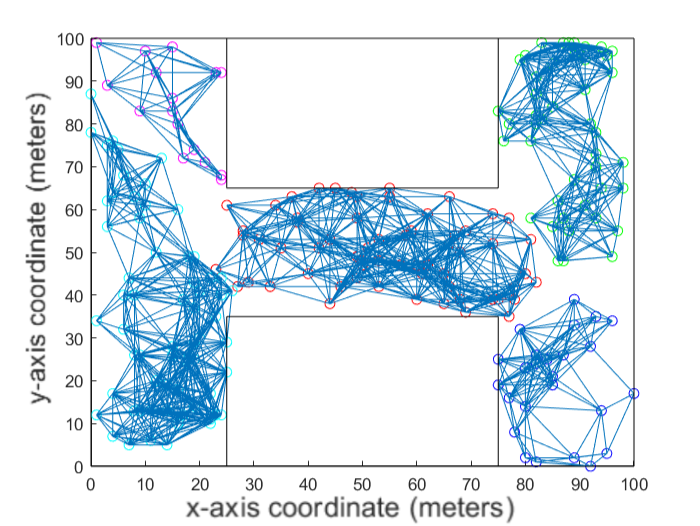}}}
		\vspace{-0.2cm}
		\caption{H-shaped obstacle}
		\label{f18}
	\end{minipage}
	\hfill
	\begin{minipage}[t]{0.49\textwidth}
		\centering
		\subfigure[]{\raisebox{-\height}{\includegraphics[scale=0.17]{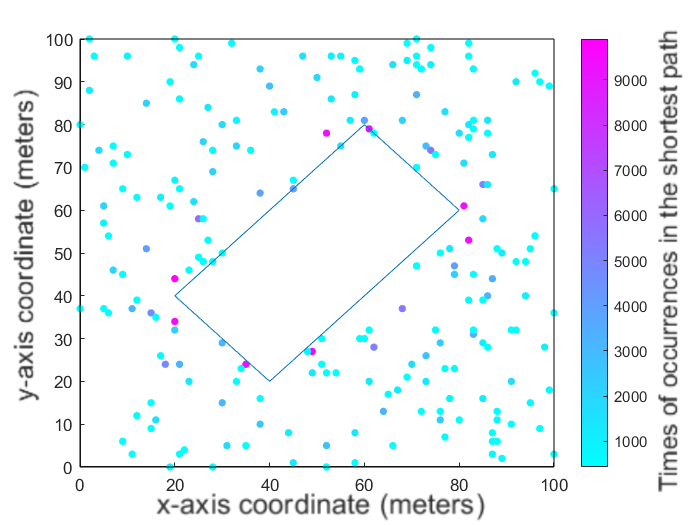}}}
		\subfigure[]{\raisebox{-\height}{\includegraphics[scale=0.17]{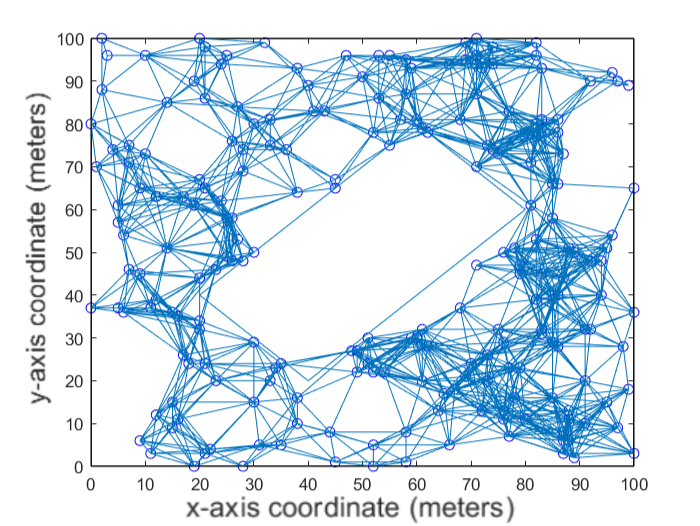}}}
		\subfigure[]{\raisebox{-\height}{\includegraphics[scale=0.17]{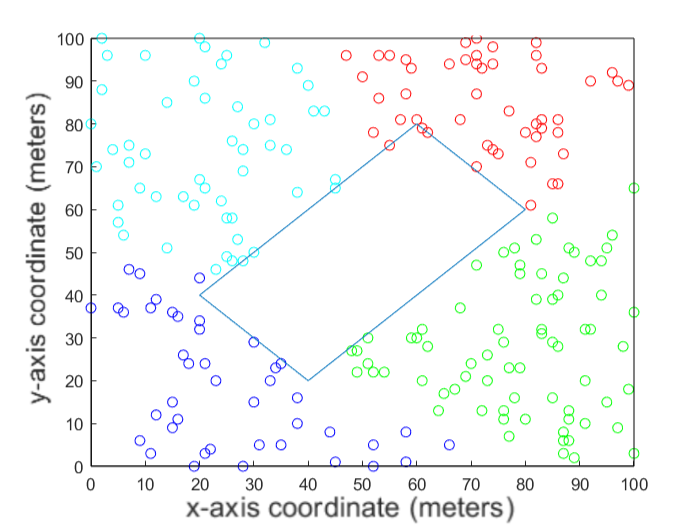}}}
		\subfigure[]{\raisebox{-\height}{\includegraphics[scale=0.17]{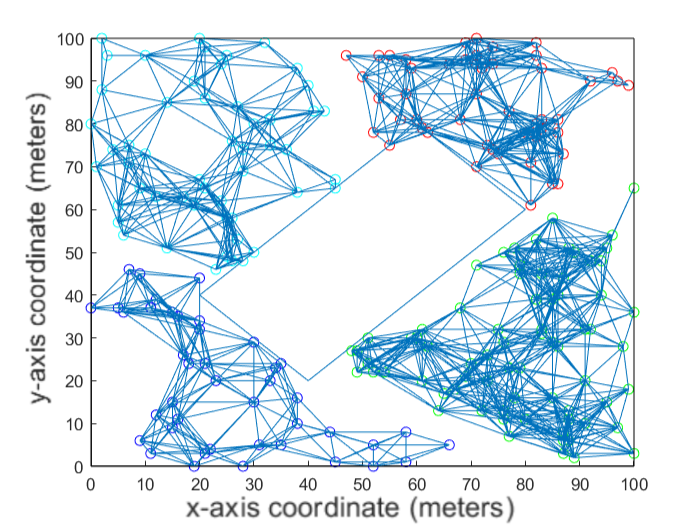}}}
		
		\vspace{-0.2cm}
		\caption{Rectangular obstacle}
		\label{f19}
	\end{minipage}
	
		\vspace{0.8cm}

	\begin{minipage}[t]{0.49\textwidth}
		\centering
		\subfigure[]{\raisebox{-\height}{\includegraphics[scale=0.17]{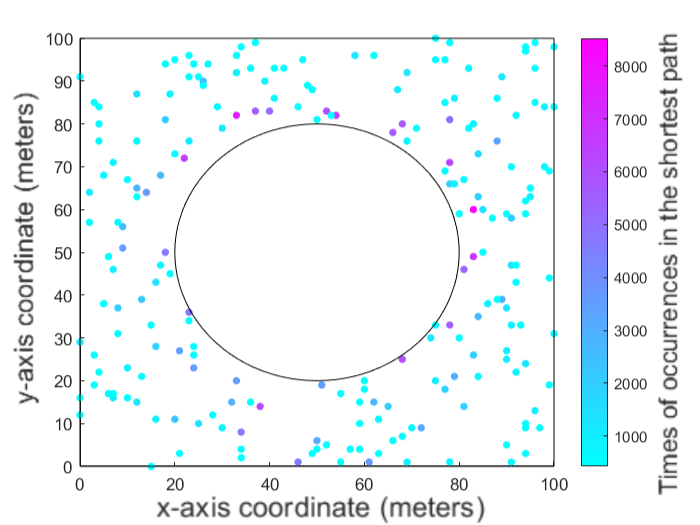}}}
		\subfigure[]{\raisebox{-\height}{\includegraphics[scale=0.17]{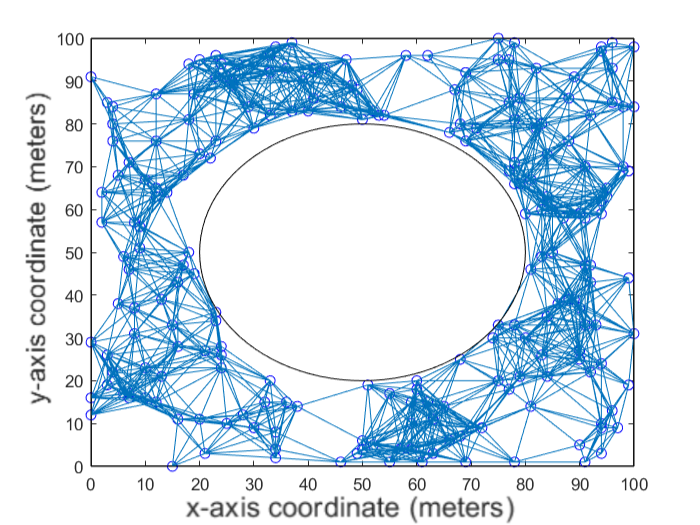}}}
		\subfigure[]{\raisebox{-\height}{\includegraphics[scale=0.17]{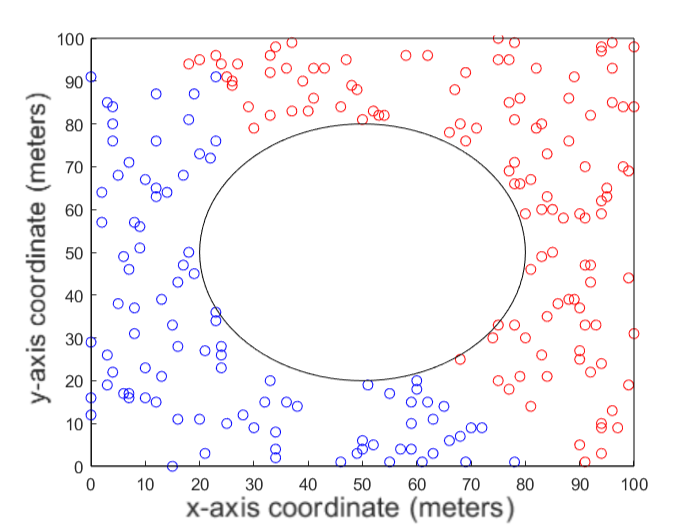}}}
		\subfigure[]{\raisebox{-\height}{\includegraphics[scale=0.17]{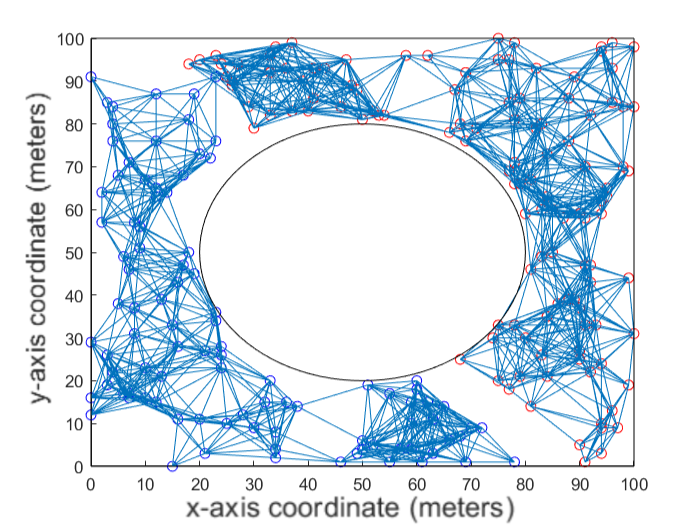}}}
		
		\vspace{-0.2cm}
		\caption{Circular obstacle}
		\label{f20}
	\end{minipage}
	\hspace{0.1cm}
	\begin{minipage}[t]{0.49\textwidth}
		\centering
		\subfigure[]{\raisebox{-\height}{\includegraphics[scale=0.17]{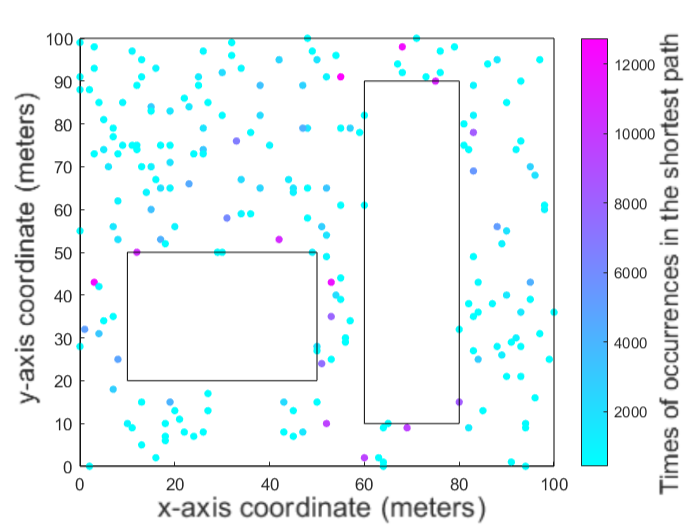}}}
		\subfigure[]{\raisebox{-\height}{\includegraphics[scale=0.17]{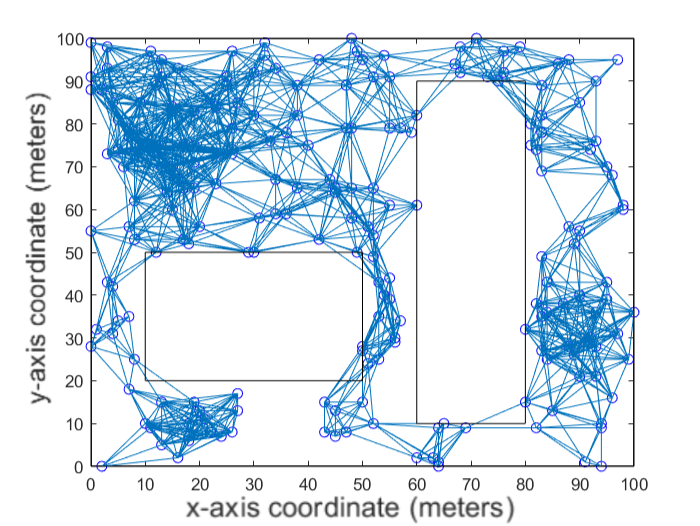}}}
		\subfigure[]{\raisebox{-\height}{\includegraphics[scale=0.17]{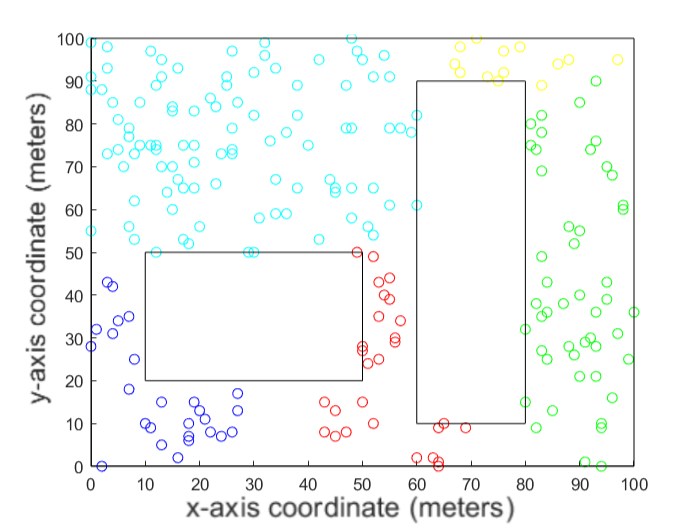}}}
		\subfigure[]{\raisebox{-\height}{\includegraphics[scale=0.17]{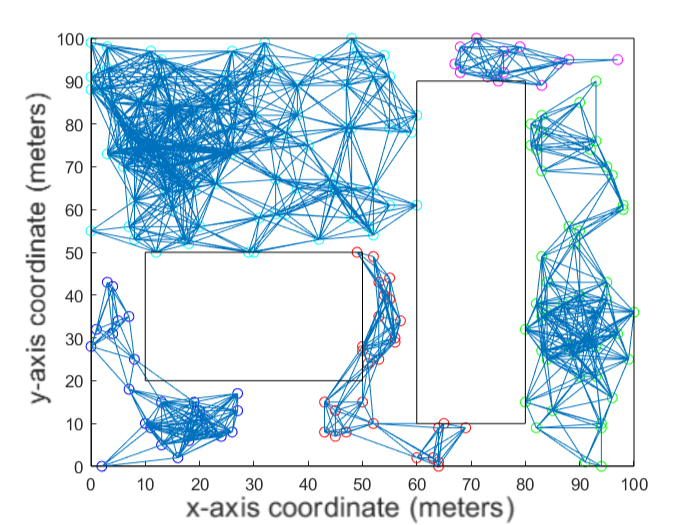}}}
		
		\vspace{-0.2cm}
		\caption{Asymmetric multi-rectangular obstacle}
		\label{f21}
	\end{minipage}
\end{figure*}

Eight different obstacle shapes are considered, including traditional forms (C-shape, S-shape, and H-shape), hole-based obstacles (rectangular and circular shapes), and complex geometric patterns specifically designed to rigorously test the partitioning method under extreme and challenging conditions (asymmetric multi-rectangular, maze-like, and smiling face shapes). The detailed spatial configurations of these obstacle layouts are illustrated in Figures \ref{f16}-\ref{f23}.

To investigate the impact of network density and node composition on partitioning and localization performance, four distinct node deployment configurations are established. These configurations feature varying ratios of unknown nodes to anchor nodes, with the unknown node ranging from 150 to 300 in increments of 50, and the corresponding number of anchor nodes ranging from 10 to 25 in increments of 5. This proportional scaling aligns with common practices in the field, as demonstrated in prior studies in the literature \cite{PHOEMPHON2020113044}. Such a setup enables a comprehensive evaluation of the model’s robustness and generalizability under different network conditions.

In each simulation scenario, all nodes are randomly distributed throughout the network while maintaining node connectivity, with the constraint that isolated nodes are not permitted, ensuring each node has at least one neighbor node within its radio communication range. Following established practices in WSNs, the maximum radio communication range for each node is set to $15 m$ \cite{article2017} to ensure reliable RSSI values for accurate distance evaluation in practical applications. To ensure statistical significance, a total of 50 independent simulations are conducted for each of the four node configurations across all obstacle scenarios. The detailed simulation parameters are summarized in Table \ref{tab2}.

\begin{table}[b]
	\caption{Simulation Parameters}
	\label{table}
	\setlength{\tabcolsep}{10pt}
	\begin{tabular}{p{95pt}|p{115pt}}
		\hline
		Parameter& Parameter Values\\
		\hline
		Network area & $100m \times 100m$ \\
		Obstacle shape & C-shape, S-shape, H-shape, rectangular, circular, asymmetric multi-rectangular, maze-like, and smiling face shapes \\
		Number of unknown nodes& $150$, $200$, $250$, $300$\\
		Number of anchor nodes& $10$, $15$, $20$, $25$\\
		Radio transmission range& $15m$\\
		\hline
	\end{tabular}
	\label{tab2}
\end{table}

\begin{table}
	\caption{Comparative analysis of theoretical convex corner counts, optimal segmentation nodes and pairs, as well as ideal partition numbers for obstacle configurations}
	\centering
	\label{table}
	\setlength{\tabcolsep}{2pt}
	\begin{tabular}{|p{37pt}!{\vrule width 1pt}p{26pt}|p{45pt}|p{55pt}|p{56pt}|}
		\hline	
		    Obstacle Shape & \makecell[c]{No. of \\ Convex \\ Corners} & \makecell[c]{Ideal No. of  \\ Segmentation \\ Nodes} & \makecell[c]{Ideal No. of \\ Segmentation \\ Node Pairs} & \makecell[c]{Ideal No. of \\ Sub-networks \\ after Partitioning} \\
		\specialrule{1pt}{0pt}{0pt}
		C-shape&    \makecell[c]{2}&\makecell[c]{4}&\makecell[c]{2}& \makecell[c]{3}\\
		\hline
		S-shape&   \makecell[c]{4}&\makecell[c]{8}&\makecell[c]{4}& \makecell[c]{5}\\
		\hline
		H-shape&   \makecell[c]{4}&\makecell[c]{8}&\makecell[c]{4}& \makecell[c]{3}\\
		\hline
		Rectangular&   \makecell[c]{4}&\makecell[c]{8}&\makecell[c]{4}& \makecell[c]{4}\\
		\hline
Circular & \makecell[c]{infinite} & \makecell[c]{$2\pi R_{c}/L$} & \makecell[c]{0, 1, 2 } & \makecell[c]{1, 2, 3} \\
		\hline
		Asymmetric multi-rectangular&\makecell[c]{8}&   \makecell[c]{16}&\makecell[c]{8}& \makecell[c]{7}\\
		\hline
		Maze-like&   \makecell[c]{6}&\makecell[c]{12}&\makecell[c]{6}& \makecell[c]{7}\\
		\hline
		Smiling face& \makecell[c]{infinite}&\makecell[c]{$2\pi R_{c}/L+8$}&\makecell[c]{4, 6, 8}& \makecell[c]{4, 5}\\
		\hline
	\end{tabular}
	\label{tab3}
\end{table}

\subsection{Analytical Optimal Partitioning Results}

To verify whether all convex features have been accurately detected, the number of segmentation node pairs is compared with the number of obstacle convex corners under various obstacle configurations. To assess whether the obstacle-induced segmentation is complete or contains redundancy, the comparison between the actual and ideal partition numbers is made, where fewer actual partitions indicate incomplete convex corner resolution and excess partitions suggest redundant segmentation. Table \ref{tab3} presents the number of convex corners, the ideal number of segmentation nodes and pairs, as well as the resulting ideal number of sub-networks formed after partitioning for the eight different obstacle scenarios, regardless of our methodology. Notably, for C-shape, S-shape and maze-like obstacles, the partitioning strictly follows the rule that the ideal number of partitions is always one greater than the ideal number of segmentation node pairs.

It is important to note that circular obstacles represent a special case in the context of ideal partitioning analysis and require separate consideration due to their unique geometric properties. Unlike polygonal obstacles with discrete convex corners, a circle is a strictly convex shape characterized by continuous curvature and the absence of abrupt angular features. As a result, signal propagation around a circular obstacle follows fundamentally different principles, that is, the link between any two nodes is either fully preserved or entirely obstructed. In such cases, signal attenuation is mainly caused by diffraction loss rather than multipath effects induced by sharp corners. Consequently, a circular obstacle can be mathematically regarded as having an infinite number of convex corners along its perimeter, leading to an ideal number of partitions that theoretically approaches infinity.

\begin{figure*}[t] 
	\centering
	\begin{minipage}[t]{0.49\textwidth}
		\centering
		\subfigure[]
		{\includegraphics[scale=0.17]{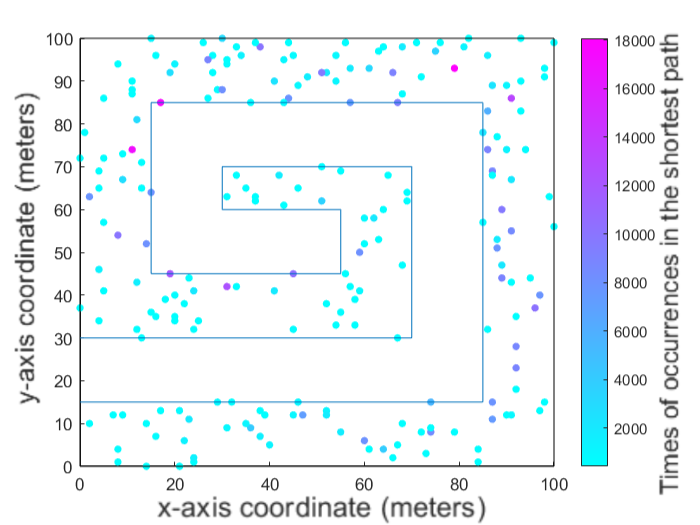}}
		\subfigure[]
		{\includegraphics[scale=0.17]{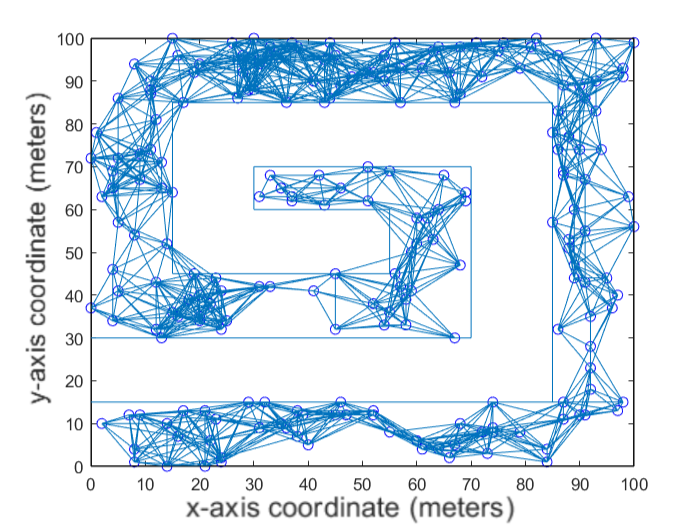}}
		\subfigure[]
		{\includegraphics[scale=0.17]{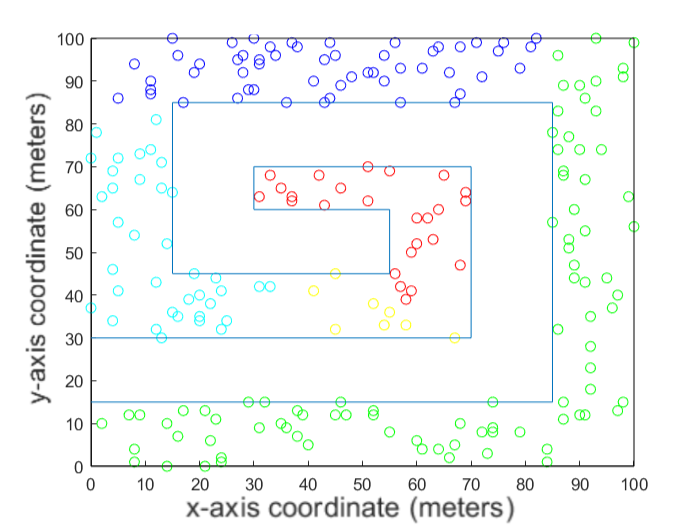}}
		\subfigure[]
		{\includegraphics[scale=0.17]{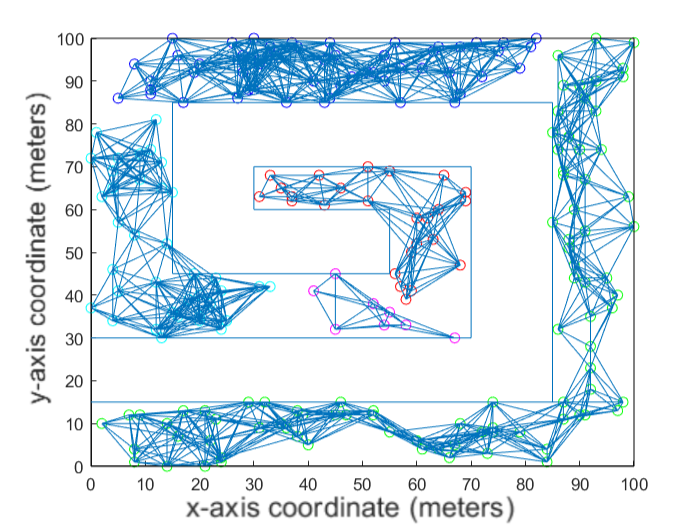}}
		\caption{Maze-like obstacle}
		\label{f22}
	\end{minipage}
	\hfill
	\begin{minipage}[t]{0.49\textwidth}
		\centering
		\subfigure[]
		{\includegraphics[scale=0.17]{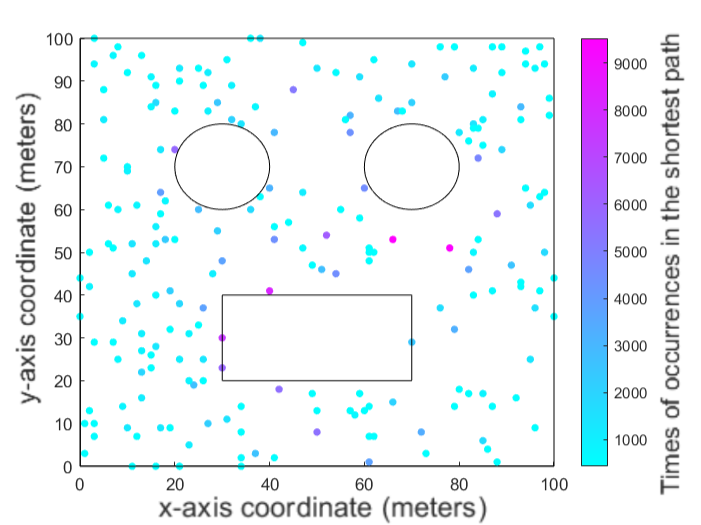}}
		\subfigure[]
		{\includegraphics[scale=0.17]{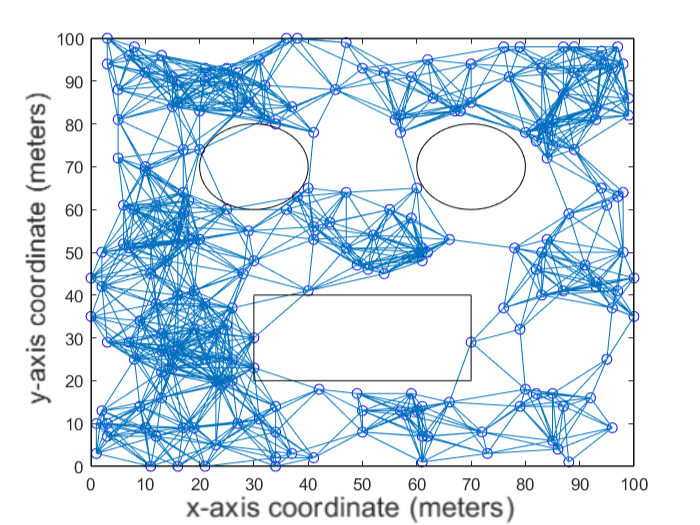}}
		\subfigure[]
		{\includegraphics[scale=0.17]{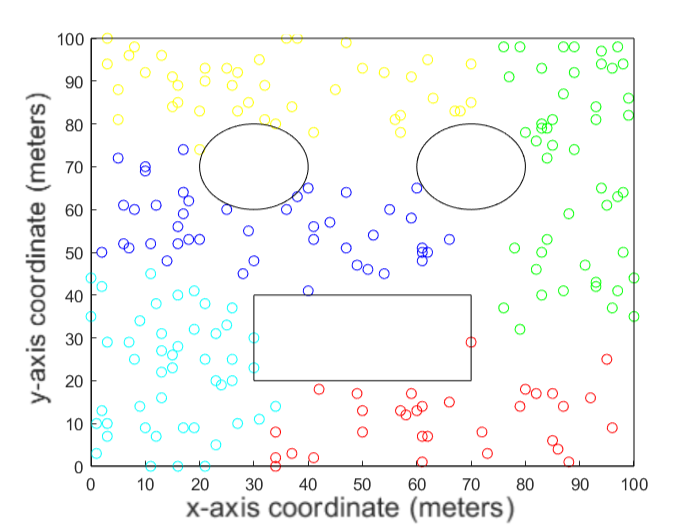}}
		\subfigure[]
		{\includegraphics[scale=0.17]{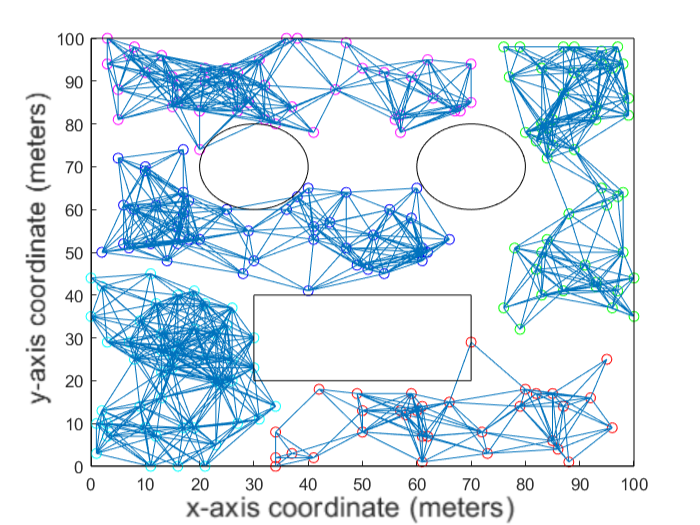}}
		\caption{Smiling face obstacle}
		\label{f23}
	\end{minipage}
\end{figure*}

However, if the partitioning is based solely on the geometric structure of the circle, it would be impractical for real-world localization applications. To address this, and taking into account the characteristics of WSNs and the communication properties of the nodes, we define the ideal number of partitions for circular obstacles as a finite value. Although the actual number of segmentation nodes required around a circular obstacle depends on node density, total node count, and spatial deployment pattern, for analytical tractability, an ideal estimate are made under the assumption that sensor nodes are uniformly distributed along the boundary of the obstacle. Under this condition, the ideal number of segmentation nodes is determined by the ratio between the circumference of the obstacle and the maximum radio transmission range of nodes, denoted by $2\pi R_{c}/L$, where $R_{c}$ is the radius of the circular object and $L=15m$. This approximation treats the boundary of the circular obstacle between adjacent nodes as a straight line rather than an arc. The use of $L$ as the denominator, rather than the average transmission range ($L/2$), is justified by the methodology adopted in this study, which emphasizes the frequency of each node's occurrence on the shortest communication path between any two nodes in the network. Consequently, the actual distance between adjacent segmentation nodes tends to approach the upper limit of the transmission range, making $L$ a more appropriate reference value.

The ideal number of segmentation node pairs can only be 0, 1, or 2 due to the geometric constraints imposed by uniform segmentation node distribution along a circular obstacle's boundary. This limited values arises from three distinct scenarios:

First, when the calculated number of segmentation nodes is less than 2 or exactly an integer larger than 2, the result may lead to either no valid segmentation pairs (if fewer than 2 segmentation nodes exist), or a fully symmetric arrangement where each node forms a pair with its neighbor, and all such pairs form a closed loop. In the latter case, nodes in every segmentation pair is also involved in another pair, causing them to effectively cancel each other out under the model's criteria for valid, non-overlapping segmentation pairs. As a result, the effective number of segmentation node pairs is 0.

Second, if the number of segmentation nodes is greater than or equal to 2 but less than 3, only 1 valid segmentation node pair can be formed. This occurs because the number of nodes is just sufficient to create a single pair, without triggering overlaps that would cancel it out.

Third, when the number of segmentation nodes is greater than 3 and non-integer, partial overlap may still occur between pairs, but due to the fractional nature of the total segmentation node count, the first and last potential pairs do not align perfectly. This prevents the formation of a complete loop of overlapping pairs, allowing two distinct and non-overlapping segmentation node pairs to exist. Therefore, under this condition, the ideal number of segmentation node pairs is 2.

Once the number of segmentation node pairs is determined, the resulting number of sub-networks after partitioning can be analytically derived. For the first two cases, where the ideal number of segmentation node pairs is 0 or 1, the number of resulting sub-networks is simply the number of pairs plus one, yielding 1 or 2 sub-networks, respectively. However, for the third case, where the ideal number of segmentation node pairs is 2, the resulting number of sub-networks may be either 2 or 3. This variation depends on whether both nodes in the second segmentation node pair fall into separate sub-networks after the initial partitioning by the first pair. If they are already in different sub-networks, no additional partitioning is required. Conversely, if both nodes remain in the same sub-network, further partitioning will be necessary, resulting in a third sub-network.


Since the smiling face obstacle also includes circular components, the ideal number of segmentation nodes, node pairs, and resulting sub-networks is analytically determined in the same manner as for circular obstacles. Specifically, the obstacle consists of two symmetric sub-circular components, each with the same diameter (e.g., $20m$ as illustrated in Figure \ref{f23}). This configuration yields an ideal segmentation node pairs of 0, 1, or 2 for each circular component. Considering the two circular components along with an additional rectangular component which has 4 convex corners, the ideal number of segmentation node pairs for the smiling face obstacle is estimated to be 4, 6, or 8. Regardless of the specific method used, the final number of partitioned sub-networks is expected to be 4 or 5, depending on whether the circular components are partitioned horizontally, vertically, or not partitioned at all.

\subsection{Accuracy Assessment Criteria for Network Partitioning and Localization Performance}

\subsubsection{Partitioning effectiveness evaluation}

To ensure accurate localization of unknown nodes, it is essential to evaluate the effectiveness of network partitioning, which aims to eliminate the impact of obstacle convex corners by dividing the network into obstacle-free sub-networks. To quantitatively assess the partitioning quality, we introduce the Average Completion Degree (ACD), a metric that reflects how well the partitioning resolves communication disruptions between nodes caused by obstacles. Specifically, ACD is defined as the proportion of node pairs whose communication is initially obstructed by obstacles but regained successful connection after partitioning. This metric directly indicates how effectively the partitioning eliminates obstacle-induced inaccuracies in distance measurements. The value of term $ACD$ ranges from 0 to 1, where a higher value signifies a more effective partitioning outcome. It can be computed as:

\begin{equation}
ACD = \frac{\sum_{i=1}^{t}(1-\frac{SPO^{after}_i}{SPO^{before}_i})}{t}
	\label{e16}
\end{equation}

In Equation \ref{e16}, parameter $t$ denotes the total number of independent  simulation trials. The metric Segment Passing through Obstacles (SPO)  quantifies the number of node-pair communication paths that intersect with obstacles in the environment. For each trial, $SPO^{before}_i$ counts the number of obstructed communication paths prior to network partitioning, while $SPO^{after}_i$ measures the remaining obstructed paths after partitioning is applied.

\subsubsection{Localization accuracy evaluation}

Regarding the localization performance, the proposed method achieves theoretically perfect localization accuracy in Cases 1 and 2 (as described in Section V-B), with the estimated node coordinates matching the ground truth positions exactly. Therefore, the evaluation primarily focuses on Case 3, where localization errors may occur. In this case, the average localization error of all unknown nodes is assessed across four different node density configurations and eight distinct obstacle geometries. 

To quantify localization accuracy, the Mean Location Error (MLE) is employed, which measures the average deviation between the estimated and actual coordinates of unknown nodes in the network. The result is expressed in meters, with lower values indicating higher localization accuracy. Term $MLE$ can be expressed as:

\begin{equation}
MLE = \frac{\sum_{i=1}^{m}\sqrt{(x_i^{true} - x_i^\mathbb{T})^2 + (y_i^{true} - y_i^\mathbb{T})^2}}{m}
\label{e17}
\end{equation}

Here, variable $m$ is the total number of unknown nodes in the network, parameter $(x_i^\mathbb{T}, y_i^\mathbb{T})$ represents the estimated coordinates of node $i$ , and term $(x_i^{true}, y_i^{true})$ denotes the corresponding ground truth coordinates in the global coordinate system.

\subsection{Evaluation of Simulated Network Partitioning}

\subsubsection{Evaluation of Simulated Segmentation Node Pairs}

Across Figures \ref{f16} to \ref{f23}, each figure consists of four sub-figures, where (a) displays the number of occurrence of each node appearing on shortest paths between any two nodes, (b) shows the pre-partitioning network connectivity status, (c) presents the partitioning results based on obstacle configurations, and (d) illustrates the post-partitioning connectivity status. These figures correspond to the case where 215 sensor nodes are randomly deployed, including 200 unknown nodes and 15 anchor nodes. During the partitioning process, the model treats all nodes equally, considering only their connectivity properties without differentiating between unknown and anchor nodes.

Table \ref{tab4} presents the simulation results, including the number of segmentation node pairs, the number of network partitions, and the ACD values for eight obstacle configurations, with node counts of 160, 215, 270, and 325. The deviations between simulated and ideal values for both the number of segmentation node pairs and number of partitions are shown in parentheses.

To compare the simulation results with theoretical expectations, the simulated number of segmentation node pairs closely corresponds to the theoretical values presented in Table \ref{tab3} for the C-shaped, S-shaped, H-shaped, and rectangular obstacles. Consequently, the number of partitioned sub-networks is consistent between the theoretical analysis and simulated outcomes, except for the case of the H-shaped obstacle.

For the H-shaped obstacle, the discrepancy arises because each bisector is generated based on the segmentation node pair located at a convex corner, resembling the angular bisector of the outer convex angle rather than following the obstacle’s edge. Consequently, even though the upper and lower rectangular components are vertically aligned, the resulting bisectors cannot merge, leading to redundant partitioning. This explains why the theoretical partition count 3 is lower than the simulated result 5, even though the number of segmentation node pairs matches the theoretical analysis.

However, notable differences are observed between the simulated results and the theoretical values in both the number of segmentation node pairs and the resulting number of partitioned sub-networks when obstacles take forms of  circular, asymmetric multi-rectangular, maze-like, or smiling face shapes.

\begin{table}[t]
	\caption{Number of convex corners, ideal number of partitions and ACD for each obstacle}
	\label{table}
	\setlength{\tabcolsep}{5pt}
	\begin{tabular}{|p{45pt}  !{\vrule width 1pt} p{35pt}|p{45pt}|p{40pt}|p{35pt}|}
		\specialrule{1pt}{0pt}{0pt}
		\makecell[c]{\makecell{Obstacle\\Shape}}& \makecell[c]{\makecell{Number\\of Sensor\\Nodes}}& \makecell[c]{\makecell{Number of\\Segmentation\\Node Pairs}}& \makecell[c]{\makecell{Number of\\Partitions}}& \makecell[c]{\makecell{$ACD$}}\\
		\specialrule{1pt}{0pt}{0pt}
		\multirow{4}{*}{C-shape}& \makecell[c]{160}& \makecell[c]{2.04(+0.04)}& \makecell[c]{3.04(+0.04)}& \makecell[c]{0.8935}\\
		& \makecell[c]{215}& \makecell[c]{2.02(+0.02)}& \makecell[c]{3.02(+0.02)}& \makecell[c]{0.9037}\\
		& \makecell[c]{270}& \makecell[c]{2(+0)}& \makecell[c]{3(+0)}& \makecell[c]{0.9249}\\
		& \makecell[c]{325}& \makecell[c]{2(+0)}& \makecell[c]{3(+0)}& \makecell[c]{0.9586}\\
		\hline
		\multirow{4}{*}{S-shape}& \makecell[c]{160}& \makecell[c]{3.36(-0.64)}& \makecell[c]{4.36(-0.64)}& \makecell[c]{0.8173}\\
		& \makecell[c]{215}& \makecell[c]{3.68(-0.32)}& \makecell[c]{4.68(-0.32)}& \makecell[c]{0.8477}\\
		& \makecell[c]{270}& \makecell[c]{3.76(-0.24)}& \makecell[c]{4.76(-0.24)}& \makecell[c]{0.8825}\\
		& \makecell[c]{325}& \makecell[c]{3.84(-0.16)}& \makecell[c]{4.84(-0.16)}& \makecell[c]{0.9218}\\
		\hline
		\multirow{4}{*}{H-shape}& \makecell[c]{160}& \makecell[c]{3.68(-0.32)}& \makecell[c]{4.48(+1.48)}& \makecell[c]{0.8698}\\
		& \makecell[c]{215}& \makecell[c]{3.92(-0.08)}& \makecell[c]{4.7(+1.7)}& \makecell[c]{0.8912}\\
		& \makecell[c]{270}& \makecell[c]{3.94(-0.06)}& \makecell[c]{4.86(+1.86)}& \makecell[c]{0.9187}\\
		& \makecell[c]{325}& \makecell[c]{3.96(-0.04)}& \makecell[c]{4.88(+1.88)}& \makecell[c]{0.9403}\\
		\hline
		\multirow{4}{*}{Rectangular}& \makecell[c]{160}& \makecell[c]{3.76(-0.24)}& \makecell[c]{3.88(-0.12)}& \makecell[c]{0.8614}\\
		& \makecell[c]{215}& \makecell[c]{3.84(-0.16)}& \makecell[c]{4(+0)}& \makecell[c]{0.8875}\\
		& \makecell[c]{270}& \makecell[c]{3.92(-0.08)}& \makecell[c]{4.2(+0.2)}& \makecell[c]{0.9098}\\
		& \makecell[c]{325}& \makecell[c]{3.94(-0.06)}& \makecell[c]{4.2(+0.2)}& \makecell[c]{0.9225}\\
		\hline
		\multirow{4}{*}{Circular}& \makecell[c]{160}& \makecell[c]{1.16(-0.84)}& \makecell[c]{2.16(+0.16)}& \makecell[c]{0.905}\\
		& \makecell[c]{215}& \makecell[c]{1.14(-0.86)}& \makecell[c]{2.14(+0.14)}& \makecell[c]{0.91}\\
		& \makecell[c]{270}& \makecell[c]{1.14(-0.86)}& \makecell[c]{2.14(+0.14)}& \makecell[c]{0.9}\\
		& \makecell[c]{325}& \makecell[c]{1.12(-0.88)}& \makecell[c]{2.12(+0.12)}& \makecell[c]{0.91}\\
		\hline
		\multirow{4}{*}{\makecell{ Asymmetric\\multi-\\rectangular}}& \makecell[c]{160}& \makecell[c]{4.04(-3.96)}& \makecell[c]{5.02(-1.98)}& \makecell[c]{0.7916}\\
		& \makecell[c]{215}& \makecell[c]{4.16(-3.84)}& \makecell[c]{5.1(-1.98)}& \makecell[c]{0.8358}\\
		& \makecell[c]{270}& \makecell[c]{4.36(-3.64)}& \makecell[c]{5.26(-1.74)}& \makecell[c]{0.8574}\\
		& \makecell[c]{325}& \makecell[c]{4.5(-3.5)}& \makecell[c]{5.36(-1.64)}& \makecell[c]{0.8784}\\
		\hline
		\multirow{4}{*}{Maze-like}& \makecell[c]{160}& \makecell[c]{4(-2)}& \makecell[c]{5(-2)}& \makecell[c]{0.8094}\\
		& \makecell[c]{215}& \makecell[c]{4.06(-1.94)}& \makecell[c]{5.06(-1.94)}& \makecell[c]{0.8228}\\
		& \makecell[c]{270}& \makecell[c]{4.16(-1.84)}& \makecell[c]{5.16(-1.84)}& \makecell[c]{0.8459}\\
		& \makecell[c]{325}& \makecell[c]{4.34(-1.66)}& \makecell[c]{5.34(-1.66)}& \makecell[c]{0.8767}\\
		\hline
		\multirow{4}{*}{Smiling face}& \makecell[c]{160}& \makecell[c]{2.56(-5.44)}& \makecell[c]{3.54(-0.46)}& \makecell[c]{0.7333}\\
		& \makecell[c]{215}& \makecell[c]{2.72(-5.28)}& \makecell[c]{3.66(-0.34)}& \makecell[c]{0.7965}\\
		& \makecell[c]{270}& \makecell[c]{3.08(-4.92)}& \makecell[c]{3.94(-0.06)}& \makecell[c]{0.8444}\\
		& \makecell[c]{325}& \makecell[c]{3.16(-4.84)}& \makecell[c]{4(+0)}& \makecell[c]{0.8997}\\
		\hline
	\end{tabular}
	\label{tab4}
\end{table}

\begin{table}[t]
	\caption{Number of convex corners and ideal number of partitions for each obstacle}
	\label{table}
	\setlength{\tabcolsep}{5pt}
	\begin{tabular}{|p{45pt}  !{\vrule width 1pt}   p{45pt}|p{35pt}|p{45pt}|p{35pt}|}
		\specialrule{1pt}{0pt}{0pt}
		\makecell[c]{\makecell{Obstacle\\Shape}}& \makecell[c]{\makecell{Number\\of Unknown\\Nodes}}& \makecell[c]{\makecell{Number\\of Anchor\\Nodes}}& \makecell[c]{\makecell{Number\\of Inaccurate \\ Nodes}}& \makecell[c]{\makecell{$MLE$\\(meters)}}\\
		\specialrule{1pt}{0pt}{0pt}
		\multirow{4}{*}{C-shape}& \makecell[c]{150}& \makecell[c]{10}& \makecell[c]{0.44}& \makecell[c]{0.0065}\\
		& \makecell[c]{200}& \makecell[c]{15}& \makecell[c]{0.02}& \makecell[c]{0.0002}\\
		& \makecell[c]{250}& \makecell[c]{20}& \makecell[c]{0}& \makecell[c]{0}\\
		& \makecell[c]{300}& \makecell[c]{25}& \makecell[c]{0}& \makecell[c]{0}\\
		\hline
		\multirow{4}{*}{S-shape}& \makecell[c]{150}& \makecell[c]{10}& \makecell[c]{0.76}& \makecell[c]{0.0228}\\
		& \makecell[c]{200}& \makecell[c]{15}& \makecell[c]{0.12}& \makecell[c]{0.0022}\\
		& \makecell[c]{250}& \makecell[c]{20}& \makecell[c]{0}& \makecell[c]{0}\\
		& \makecell[c]{300}& \makecell[c]{25}& \makecell[c]{0}& \makecell[c]{0}\\
		\hline
		\multirow{4}{*}{H-shape}& \makecell[c]{150}& \makecell[c]{10}& \makecell[c]{0.58}& \makecell[c]{0.0182}\\
		& \makecell[c]{200}& \makecell[c]{15}& \makecell[c]{0}& \makecell[c]{0}\\
		& \makecell[c]{250}& \makecell[c]{20}& \makecell[c]{0}& \makecell[c]{0}\\
		& \makecell[c]{300}& \makecell[c]{25}& \makecell[c]{0}& \makecell[c]{0}\\
		\hline
		\multirow{4}{*}{Rectangular}& \makecell[c]{150}& \makecell[c]{10}& \makecell[c]{0.3}& \makecell[c]{0.0071}\\
		& \makecell[c]{200}& \makecell[c]{15}& \makecell[c]{0}& \makecell[c]{0}\\
		& \makecell[c]{250}& \makecell[c]{20}& \makecell[c]{0}& \makecell[c]{0}\\
		& \makecell[c]{300}& \makecell[c]{25}& \makecell[c]{0}& \makecell[c]{0}\\
		\hline
		\multirow{4}{*}{Circular}& \makecell[c]{150}& \makecell[c]{10}& \makecell[c]{0.06}& \makecell[c]{0.0065}\\
		& \makecell[c]{200}& \makecell[c]{15}& \makecell[c]{0}& \makecell[c]{0}\\
		& \makecell[c]{250}& \makecell[c]{20}& \makecell[c]{0}& \makecell[c]{0}\\
		& \makecell[c]{300}& \makecell[c]{25}& \makecell[c]{0}& \makecell[c]{0}\\
		\hline
		\multirow{4}{*}{\makecell{ Asymmetric\\multi-\\rectangular}}& \makecell[c]{150}& \makecell[c]{10}& \makecell[c]{0.68}& \makecell[c]{0.0142}\\
		& \makecell[c]{200}& \makecell[c]{15}& \makecell[c]{0.02}& \makecell[c]{0.0002}\\
		& \makecell[c]{250}& \makecell[c]{20}& \makecell[c]{0}& \makecell[c]{0}\\
		& \makecell[c]{300}& \makecell[c]{25}& \makecell[c]{0}& \makecell[c]{0}\\
		\hline
		\multirow{4}{*}{Maze-like}& \makecell[c]{150}& \makecell[c]{10}& \makecell[c]{0.68}& \makecell[c]{0.0279}\\
		& \makecell[c]{200}& \makecell[c]{15}& \makecell[c]{0.28}& \makecell[c]{0.0085}\\
		& \makecell[c]{250}& \makecell[c]{20}& \makecell[c]{0.06}& \makecell[c]{0.0005}\\
		& \makecell[c]{300}& \makecell[c]{25}& \makecell[c]{0}& \makecell[c]{0}\\
		\hline
		\multirow{4}{*}{Smiling face}& \makecell[c]{150}& \makecell[c]{10}& \makecell[c]{0.22}& \makecell[c]{0.0059}\\
		& \makecell[c]{200}& \makecell[c]{15}& \makecell[c]{0}& \makecell[c]{0}\\
		& \makecell[c]{250}& \makecell[c]{20}& \makecell[c]{0}& \makecell[c]{0}\\
		& \makecell[c]{300}& \makecell[c]{25}& \makecell[c]{0}& \makecell[c]{0}\\
		\hline
	\end{tabular}
	\label{tab5}
\end{table}

In the case of the circular obstacle with a diameter of $60 m$, the theoretical analysis suggests that there should be 2 segmentation node pairs, resulting in 2 or 3 sub-networks after partitioning. However, the simulation results indicate the presence of only 1 segmentation node pair on average, leading to 2 partitions. This discrepancy arises from the fact that, unlike the theoretical analysis where nodes are assumed to be uniformly distributed with segmentation nodes positioned along the boundary of the circle, in the simulation, nodes are randomly deployed throughout the network and may not be located near the circular boundary. This randomness can lead to either an increase or a decrease in the actual number of segmentation node pairs. Therefore, the number of segmentation node pairs may vary between 0, 1, and 2, depending on whether potential pairs neutralize each other, nodes are beyond one-hop communication range except for a single valid pair, or only the initial and terminal pairs on the loop remain uncanceled.

To assess whether the simulated number of partitions accurately reflects the degree to which the obstacle impedes connectivity, we conducted additional simulations to evaluate the average number of node pairs that traverse the circular obstacle per simulation. The results are presented in Table \ref{tab6}, where circular obstacles are with diameters of $15m$, $30m$, $45m$, $60m$, $75m$, and $90m$. For each diameter, the total number of nodes is set to 160, 215, 270, and 325, respectively. For each combination of obstacle diameter and node count, the simulation runs 500 times to ensure statistical robustness.

It can be observed that when the diameter is $60 m$ and the number of nodes increases from 160 to 325, the average number of node pairs that traverse the circular obstacle aligns with the value in Table \ref{tab6}, though slightly higher. This is because not all traversing node pairs necessarily constitute segmentation node pairs. Therefore, the number of traversing node pairs is expected to be equal to or greater than the number of segmentation node pairs.

As the number of nodes increases, the average number of node pairs spanning the obstacle increases accordingly. This is because higher node density results in more nodes being positioned near the boundary of the circular obstacle, thereby increasing the likelihood of forming communication links across the obstacle. Additionally, it can be observed that for each node count, the average number of traversing node pairs is minimum when the obstacle diameter is $45m$. This optimal diameter emerges from a fundamental trade-off between the obstacle's geometric properties. When the diameter falls below $45 m$, the resulting higher curvature weakens links not only between nodes adjacent to the obstacle but also those nearby, as more signal paths intersect the obstacle. Conversely, when the diameter exceeds $45 m$, the reduced deployment area outside the obstacle forces denser node placement, increasing the probability of nodes being adjacent to the boundary of the circular obstacle.

The simulation statistics presented in Table \ref{tab7} indicate that the average ratio of node communications crossing the circular obstacle is primarily determined by the obstacle diameter. This is because increasing the number of nodes results in not only more paths crossing the obstacle, but also more paths not crossing it, keeping the overall traversal ratio relatively constant and consistently low across different network densities. These findings validate the data in Table \ref{tab4}, demonstrating strong consistency between the simulated and actual number of segmentation node pairs, and further confirming the appropriateness of the theoretical values set for circular obstacle in Table \ref{tab3}.



\begin{table}[htbp]
	\centering
	\caption{Average number of node pairs traversing circular obstacle per simulation over 500 runs}
	\label{tab:experiment}
	\begin{tabular}{l *{6}{S[table-format=1.3]}}
		\toprule
		\multirow{2}{*}{No. of nodes} & \multicolumn{6}{c}{Diameter of circular obstacle (m)} \\
		\cmidrule(lr){2-7}
		& 15 & 30 & 45 & 60 & 75 & 90 \\
		\midrule
		160  & 1.524 & 0.704 & 0.558 & 0.696 & 0.898 & 1.210 \\
		215  & 2.718 & 1.306 & 0.932 & 1.240 & 1.512 & 2.096 \\
		270  & 4.320 & 1.984 & 1.426 & 1.826 & 2.396 & 3.456 \\
		325  & 6.706 & 2.828 & 2.060 & 2.768 & 3.566 & 4.896 \\
		\bottomrule
	\end{tabular}
	\label{tab6}
\end{table}

\begin{table}[htbp]
	\centering
	\caption{Average ratio of node communication traversing circular obstacle (\%)}
	\label{tab:performance}
	\setlength{\tabcolsep}{5pt}
	\begin{tabular}{cccccccc}
		
		\toprule
		\multirow{2}{*}{No. of nodes} & \multicolumn{6}{c}{Diameter of circular obstacle (m)} \\
		\cmidrule{2-7}
		& 15 & 30 & 45 & 60 & 75 & 90 \\
		\midrule
		160 & 0.1991 & 0.0896 & 0.0666 & 0.0769 & 0.0888 & 0.1035 \\
		215 & 0.1944 & 0.0912 & 0.0616 & 0.0751 & 0.0825 & 0.0990 \\
		270 & 0.1957 & 0.0876 & 0.0598 & 0.0701 & 0.0826 & 0.1032 \\
		325 & 0.2094 & 0.0860 & 0.0593 & 0.0733 & 0.0846 & 0.1012 \\
		\midrule
		Average & 0.1997 & 0.0886 & 0.0618 & 0.0739 & 0.0846 & 0.1017 \\
		\bottomrule
	\end{tabular}
	\label{tab7}
\end{table}

For the asymmetric multi-rectangular and maze-like obstacles, the discrepancies between the simulated and theoretical results for the number of segmentation node pairs and sub-network partitions can be attributed to two main factors. In scenarios where obstacles are positioned near the network boundary, the reduced node density around certain convex corners limits the K-means algorithm's ability to reliably identify these features as segmentation nodes. Additionally, geometric configurations where candidate segmentation nodes become enclosed by existing node pairs further contribute to the underestimation. Together, these factors typically lead to 2 to 4 fewer segmentation node pairs and 2 fewer sub-network partitions in simulations compared to theoretical expectations.

For the smiling face obstacle, the discrepancies between the simulated and theoretical results arise from the combined effects of curvature-dependent phenomena of circular obstacles and boundary occlusion issues associated with complex polygonal structures like asymmetric multi-rectangular and maze-like obstacles.

\subsubsection{Evaluation of ACD Simulation Results}

\begin{figure*}[hpbt] 
	\centering
	\begin{minipage}[t]{0.48\textwidth}
		\centering
		\includegraphics[width=\linewidth]{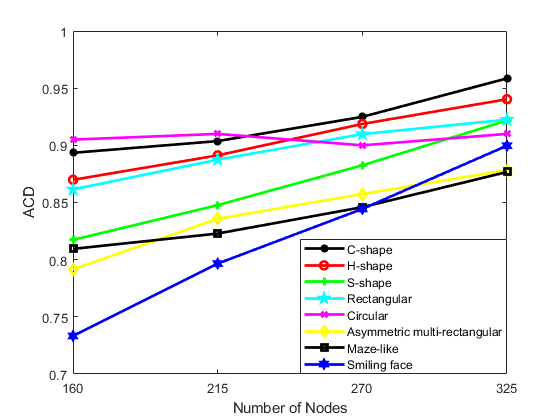}
		\caption{The ACD values of self-organizing partitions across various obstacle shapes and node quantities.}
		\label{f24}
	\end{minipage}
	\hfill 
	\begin{minipage}[t]{0.48\textwidth}
		\centering
		\includegraphics[width=\linewidth]{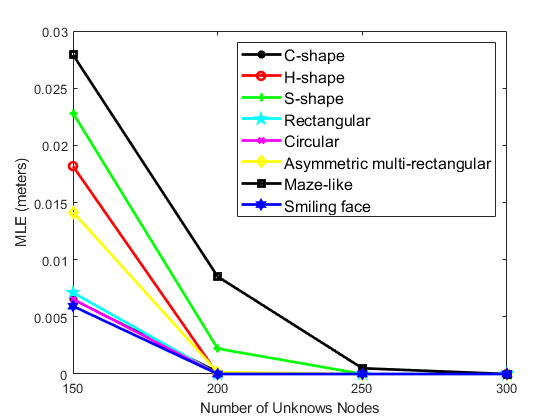}
		\caption{Mean location error of self-organizing partitions across various obstacle shapes and node quantities.}
		\label{f25}
	\end{minipage}
\end{figure*}



To visualize the ACD values of all the obstacles listed in Table \ref{tab4} and analyze their trends with respect to node density, Figure \ref{f24} presents the ACD measurements for the eight obstacle configurations across the four different node densities. Each ACD value is derived from 50 simulations per node density. The overall average value of ACD across all obstacle configurations and node densities is 0.874, which indicates consistently satisfactory partitioning results, as this value approaches the ideal benchmark of 1.

It can be observed that the sensitivity of ACD to node density varies across different obstacle types, with the smiling face obstacle exhibiting the highest sensitivity due to its complex geometry, which combines asymmetric circular and polygonal structures. This complexity together with the random distribution of nodes causes substantial disparities in communication interactions near convex corners. Under low node density conditions, these geometric features often lead to undetected corners and incomplete partitioning. However, increasing node density enables more accurate detection of convex corners and produces non-redundant partitions, as a sufficient number of nodes better capture the obstacle's complete connectivity profile. Consequently, the smiling face obstacle shows the greatest ACD improvement across incremental density levels.

The partitioning performance of ACD for maze-like and asymmetric multi-rectangular obstacles is illustrated in Figure \ref{f24}. These obstacles are designed to test network partitioning performance in narrow and elongated regions. Unlike the smiling face obstacle, the ACD value shows only a small increase as node density rises, and partitioning performance remains almost consistent across different densities, with the ACD value varying by no more than 0.09. This stability arises from the limited node deployment area and the close proximity between obstacle edges or between an obstacle edge and the network boundary. Additionally, sparse connectivity of nodes near certain convex corners hinders the formation of effective segmentation node pairs, preventing partitions in these regions. The extreme narrowness of the segmentation area can also induce partition shifts, further reducing the ACD. While higher node density mitigates these issues to some degree, these obstacles still achieve the lowest ACD values among all tested obstacle types, particularly at high node densities.

As shown in Table \ref{tab4}, for traditional obstacle shapes including C-shaped,  H-shaped, S-shaped, and rectangular obstacles, both the number of segmentation node pairs and partitions closely approximates their ideal values, yielding consistently high ACD values above 0.92.

Specifically, for the C-shape obstacle, when deploying 160 and 215 nodes, the average number of partitions slightly exceeds the ideal value of 3, reaching 3.04 and 3.02, indicating partition redundancy. This phenomenon stems from the completely random node deployment, which can lead to uneven node distribution. In extreme cases, a node-free hole may emerge within the deployment area. When the minimum radial span of such a hole exceeds the maximum radio communication range of nodes, direct communication across the hole becomes impossible, forcing multi-hop routing around it. This significantly increases the occurrence of nodes surrounding the hole on shortest paths, potentially causing their erroneous identification as segmentation nodes. Consequently, excessive segmentation node pairs form, creating redundant partitions. However, higher node densities substantially reduce the probability of such redundancy. Among all obstacle types, the C-shape achieves the highest ACD value of 0.96.

For the H-shape obstacle, ideal partitioning should divide the network into three distinct areas. However, our simulation results show that the number of segmentation node pairs is slightly fewer than the convex corners. This occurs because at lower node densities, the node distribution becomes uneven, and the symmetric geometry enables multiple shortest paths between most pairs of nodes. For example, the top-left nodes can route through multiple intermediate nodes in the central area between the two rectangular components to reach either top-right or bottom-right nodes via equally optimal paths. Consequently, nodes near convex corners don't consistently dominate the shortest paths, making them less distinguishable from other nodes for segmentation purposes. As node density increases, this issue is progressively mitigated. However, partition redundancy persists across all densities because the algorithm can only perform bisector-based partitioning at convex corners. Nevertheless, this redundancy doesn't affect partitioning performance, as evidenced by the high ACD values.

For the rectangular obstacle, although the simulated number of segmentation node pairs and resulting partitions closely aligns with theoretical expectations, the ACD value remains at 0.92 and does not reach 1, even under high node densities. This discrepancy arises because the bisectors generated by our algorithm do not originate precisely from the exact convex corners, as the obstacle’s shape is unknown to nodes. As a result, even after partitioning, some node communication may still traverse the convex corners of the obstacle. This fundamental limitation persists regardless of network density, explaining why the ACD value has a limited performance gain from 0.91 to 0.92, when node count rises from 270 to 325.

The S-shaped obstacle exhibits intermediate geometric complexity between the simple rectangular obstacle and the more intricate maze-like and asymmetric multi-rectangular obstacles. Consequently, the observed ACD deviation from theoretical values can be explained by a combination of the factors affecting all three obstacle types. This hybrid nature is reflected in the S-shape's ACD values, which consistently fall between those measured for the three obstacle types.

The circular obstacle presents a special case where most communication paths between nodes do not intersect it. As shown in Table \ref{tab6}, when the obstacle radius is $30m$, the average number of node pairs traversing the obstacle increases only slightly, from 0.696 to 2.768 as the total node count rises from 160 to 325. This indicates that the growth in traversing node pairs remains gradual as network density changes. Additionally, the value of $SPO^{after}$ scales nearly proportionally with $SPO^{before}$ in Equation \ref{e16}, yielding the ACD value consistently falls between 0.9 and 0.91, as shown in Table \ref{tab4}. The ACD value does not reach the ideal value of 1 is due to two reasons. First, because the number of traversing node pairs is either equal to or smaller than the number of segmentation node pairs, a result of the algorithm’s built-in redundancy elimination mechanism.  Second, although nodes near the obstacle appear more frequently on shortest paths, their random spatial distribution creates variability in their appearing frequencies. Thus, not all nodes near the boundary are selected as segmentation nodes, and partitioning only applies to a subset of traversing node pairs rather than all of them.

\subsection{Localization Accuracy of Unknown Nodes}

The localization accuracy of unknown nodes in the presence of unknown obstacles is dependent on network partitioning performance. As shown in Table \ref{tab5}, when the total number of nodes is 160 or 215, the MLE value is non-zero across nearly all obstacle configurations, with the highest observed MLE reaching 0.0279. In these cases, the proportion of inaccurately positioned nodes ranges between $0\%$ and $0.5\%$. However, as the node count increases to 270, the MLE converges to 0, demonstrating that our algorithm achieves perfect localization accuracy under relatively dense network conditions. Across eight obstacle configurations and four node densities, the average MLE is only 0.0038, confirming that our method achieves $>99.9\%$ localization accuracy even in large-scale networks containing unknown obstacles. This performance holds under the condition of accurate RSSI-based distance estimation.

The localization inaccuracy under low network densities stems from five factors. Firstly, network partitioning reduces the number of unknown nodes in each sub-network, significantly increasing the probability of Case 3, as detailed in Section V.B. In this scenario, an unknown node may have only one reference node, forcing its estimated position to be computed as the median point of the exclusive curve derived from inaccessible surrounding nodes. However, as illustrated in Figure \ref{f15}, the actual node position could lie anywhere along this curve, introducing inherent positioning errors when using the median-point approximation.

Secondly, the self-organized node placement affects bisector determination. Since nodes rely solely on connectivity to establish bisectors, these bisectors may not originate precisely from obstacle convex corners, particularly when node distribution is non-uniform. Consequently, certain convex corners may remain unaddressed in some partitioned sub-networks. When there are unknown nodes positioning adjacent to such unprocessed corners, their communication links become susceptible to obstruction-induced errors, leading to inaccurate distance measurements.

Thirdly, the algorithm's redundancy elimination mechanism may prevent the use of certain valid segmentation node pairs, particularly in narrow or elongated network regions. In such cases, an excessive number of candidate segmentation pairs can lead to mutual cancellation during the selection process. This  explains the observed peak in the number of inaccurate nodes for S-shaped, maze-like, and asymmetric multi-rectangular obstacles under low network density conditions, where the cancellation effect is most pronounced. Nevertheless, this inaccuracy only has limited impact on the MLE results, as higher node density within sub-networks reduces the likelihood of Case 3 scenario occurring.

Fourthly, the identified segmentation node pairs may not always be optimal due to limitations in the shortest-path occurrence metric. This inaccuracy arises because nodes appearing most frequently on shortest paths are not necessarily the most appropriate segmentation points. The selection is influenced by the obstacle's inherent geometry, network density, and stochastic node distribution. For symmetric obstacle geometries like H-shaped obstacle, this sub-optimal selection may occur and lead to a higher MLE of 0.58 compared to other obstacle types.

Finally, errors can arise during the transformation from the relative coordinate system to the globe coordinate system. This is because, in each partitioned sub-network, the relative coordinates of all nodes are determined based on their connectivity with neighboring nodes. Since each sub-region contains only three anchor nodes, it is possible that one, two, or even all of these anchor nodes have their relative positions estimated through Case 3 scenario. As a result, when converting from the relative to the globe coordinate system, the correction matrix, as shown in Equation \ref{e15}, including factors $R_{1}$, $R_2$, $R_3$, $R_4$, $\Delta x$, and $\Delta y$, that rely on these anchor nodes introduce errors, leading to inaccuracies in the absolute positions of unknown nodes.



\section{Conclusion}

Accurate indoor localization in WSNs remains a critical challenge due to signal distortions caused by obstacles in complex indoor environments. To address this challenge, this paper proposed a robust localization algorithm that enhances positioning accuracy by strategically identifying and severing obstructed communication links while leveraging network topology. The key contributions involve obstacle-aware distance evaluation through analyzing RSSI-based measurement errors induced by various obstacle configurations, including convex and concave corners. The method identifies critical segmentation nodes along obstructed paths to partition the network into obstacle-free sub-regions, then transforms convex obstacles into concave configurations via bisector establishment for precise sub-area localization. By implementing relative coordinate systems for each sub-network and subsequently calibrating them into a global framework, the algorithm achieves exceptional Localization accuracy exceeding $99.99\%$. Simulation demonstrated the approach's effectiveness, with successful severing of $87\%$ of obstacle-affected paths across diverse obstacle shapes and node densities, confirming its practical applicability in real-world indoor localization scenarios.



\ifCLASSOPTIONcaptionsoff
  \newpage
\fi



%




\bibliographystyle{IEEEtran}
\bibliography{reference}

%








\end{document}